\documentclass[prd,preprint,superscriptaddress,tightenlines,nofootinbib,eqsecnum,showpacs]{revtex4}

\usepackage{bm}
\usepackage{amsmath}
\usepackage{amsfonts}
\usepackage{amssymb}
\usepackage{hyperref}
\usepackage{mathrsfs}
\usepackage{graphicx}

\allowdisplaybreaks
\DeclareSymbolFontAlphabet{\mathrsfs}{rsfs}
\DeclareMathAlphabet{\mathcal}{OMS}{cmsy}{m}{n}

\newcommand{\scri}{\mathrsfs{I}}
\newcommand{\ud}{\mathrm{d}}
\newcommand{\beq}{\begin{equation}}
\newcommand{\eeq}{\end{equation}}

\newcommand{\rdot}{{\dot{\cal R}}_1}
\newcommand{\ubar}{{\bar u}}
\newcommand{\rddot}{{\ddot{\cal R}}_1}

\begin{document}

\title{The First Law of Binary Black Hole Mechanics \\ in General Relativity and Post-Newtonian Theory}

\author{Alexandre Le Tiec}\email{letiec@umd.edu}
\affiliation{Maryland Center for Fundamental Physics \& Joint Space-Science Institute, Department of Physics, University of Maryland, College Park, MD 20742, USA}

\author{Luc Blanchet}\email{blanchet@iap.fr}
\affiliation{$\mathcal{G}\mathbb{R}\varepsilon{\mathbb{C}}\mathcal{O}$, Institut d'Astrophysique de Paris --- UMR 7095 du CNRS, \\ Universit\'e Pierre \& Marie Curie, 98\textsuperscript{bis} boulevard Arago, 75014 Paris, France}

\author{Bernard F. Whiting}\email{bernard@phys.ufl.edu}
\affiliation{Institute for Fundamental Theory, Department of Physics, University of Florida, Gainesville, FL 32611, USA}

\date{\today}

\begin{abstract}
First laws of black hole mechanics, or thermodynamics, come in a variety of different forms. In this paper, from a purely post-Newtonian (PN) analysis, we obtain a first law for binary systems of point masses moving along an exactly circular orbit. Our calculation is valid through 3PN order and includes, in addition, the contributions of logarithmic terms at 4PN and 5PN orders. This first law of binary point-particle mechanics is then derived from first principles in general relativity, and analogies are drawn with the single and binary black hole cases. Some consequences of the first law are explored for PN spacetimes. As one such consequence, a simple relation between the PN binding energy of the binary system and Detweiler's redshift observable is established. Through it, we are able to determine with high precision the numerical values of some previously unknown high order PN coefficients in the circular-orbit binding energy. Finally, we propose new gauge invariant notions for the energy and angular momentum of a particle in a binary system.
\end{abstract}

\pacs{04.25.Nx, 04.30.-w, 04.80.Nn, 97.60.Jd, 97.60.Lf}

\maketitle

\section{Introduction}\label{sec:introduction}

\subsection{Motivation}\label{subsec:motivation}

Inspiralling and coalescing binary systems composed of black holes and/or neutron stars are among the most promising sources of gravitational radiation that might be detectable by forthcoming ground-based interferometers, such as Advanced LIGO and Advanced Virgo, as well as by future space-based observatories~\cite{CuTh.02}. The detection and analysis of these signals require very accurate theoretical predictions, for use in the construction of gravitational-wave templates. Given the complexity of the Einstein field equations, continued progress requires that we must (i) cleverly identify which parts of the problem naturally lend themselves to simplification, and (ii) rely on a combination of approximation and numerical methods.

There are two main approximation schemes for studying the relativistic dynamics of compact binary systems, and the associated emission of gravitational radiation: the post-Newtonian (PN) approximation, which is well-suited to describe the inspiralling phase of arbitrary mass ratio compact binaries in the small velocity and weak field regime ($v \ll c$)~\cite{Bl.06}, and black hole perturbation theory, which provides an accurate description of extreme mass ratio binaries ($m_1 \ll m_2$), even in the strong field regime~\cite{Po.al.11}.

Each of these approximation methods frequently relies on a simplified description of one or both of the compact objects in terms of structureless \textit{point particles}, characterized solely by their masses $m_1$ and $m_2$, and eventually their spins. Even though the notion of a point mass has been shown to be ill-defined in the exact theory of general relativity~\cite{GeTr.87,Wa.11},\footnote{Physically, if one tries to compress an extended body down to a single point, a black hole will form before the point-particle limit is reached.} it can be made sensible in approximation methods such as PN expansions or black hole perturbation theory. This idealization is particularly convenient in order to carry out calculations up to very high orders, as required for gravitational-wave searches relying on matched filtering.

Except for the occurrence of a gradual inspiral driven by radiation-reaction, the orbits of stellar mass compact binaries can be considered to be circular, to a very high degree of approximation. Mathematically, the approximation of an exactly closed circular orbit translates into the existence of a \textit{helical Killing vector} (HKV), along the orbits of which the spacetime geometry is invariant. This HKV field (say $K^\alpha\partial_\alpha = \partial_t + \Omega \, \partial_\varphi$, where $\Omega$ is the constant circular orbit frequency) can be viewed as the generator of time translations in a co-rotating frame. Given the high astrophysical relevance of this approximation, helically symmetric spacetimes have been studied extensively in the literature~\cite{De.89,BlDe.92,De.94,Go.al.02,Gr.al.02,Fr.al.02,Kl.04,Be.al.06,FrUr.06,Be.al2.07,HePr.09}; these numerous works range from formal mathematical analyses to practical calculations of initial data for binary black holes and binary neutron stars.

In general relativity, it is known that helically symmetric spacetimes cannot be asymptotically flat~\cite{Kl.04}. This can easily be understood from a heuristic point of view: in order to maintain the binary on a fixed circular orbit, the energy radiated in the form of gravitational waves needs to be compensated by an equal amount of incoming radiation. Far away from the source, the resulting system of standing waves rapidly dominates the energy content of the spacetime, such that the falloff conditions necessary to ensure asymptotic flatness cannot be satisfied. Nevertheless, asymptotic flatness can be recovered if, loosely speaking, the gravitational radiation can be ``turned off''. There are two well-known approximations to general relativity for which this can be achieved: the Isenberg, Wilson and Mathews approximation (or conformal flatness condition)~\cite{IsNe.80,Is.08,WiMa.89}, and the post-Newtonian approximation. In the latter, it is in principle possible to unambiguously disentangle the \textit{conservative} part of the orbital dynamics, from the \textit{dissipative} effects related to radiation-reaction (at least up to 3.5PN order\footnote{As usual we refer to $n$PN as the order corresponding to terms $\mathcal{O}(c^{-2n})$ in the equations of motion.}). Consequently, in the PN approximation, one can justifiably consider non-radiative, helically symmetric, asymptotically flat spacetimes. 

\subsection{Overview}\label{subsec:overview}

In this paper we shall consider such non-radiative, helically symmetric, asymptotically flat spacetimes, in which both compact objects are modelled as point masses $m_1$ and $m_2$ moving on exact circular orbits. Building on previous works~\cite{Bl.al.10,Bl.al2.10},\footnote{The papers~\cite{Bl.al.10} and~\cite{Bl.al2.10} are hereafter referred to as Papers I and II, respectively.} we compute Detweiler's redshift observables $z_1$ and $z_2$~\cite{De.08}, which represent the redshift of light rays emitted from the particles and received on the helical symmetry axis perpendicular to the orbital plane. On the other hand we compute also the total ADM mass $M$ and total angular momentum $J$ of the system of two point particles. Our calculations are carried through high post-Newtonian order, being complete up to 3PN order and including also the leading-order 4PN and next-to-leading 5PN logarithmic contributions. 

The ADM mass $M$, angular momentum $J$, and redshifts $z_{1,2}$ are all functions of the three independant variables of the problem, namely the orbital frequency $\Omega$ that is imposed by the existence of the HKV, and the individual masses $m_1$ and $m_2$ of the particles. We shall prove from our high-order post-Newtonian calculation that the variations of the ADM quantities are linked to the variations of the individual masses by the first law of the binary point-particle mechanics (or ``thermodynamics'')
\beq\label{first_law0}
	\delta M - \Omega \, \delta J = z_1 \, \delta m_1 + z_2 \, \delta m_2 \,.
\eeq
We shall demonstrate that this law is actually a particular case, valid when one assumes the existence of a HKV covering the entire spacetime, of the generalized law of black hole mechanics obtained by Friedman, Ury{\=u}, and Shibata~\cite{Fr.al.02}. Various consequences of this law will also be investigated and discussed in the framework of post-Newtonian spacetimes, notably the interesting relation
\beq\label{first_integral0}
	M - 2 \Omega J = m_1 z_1 + m_2 z_2 \, .
\eeq
As an application of the first law \eqref{first_law0} we shall be able to determine the numerical values of some previously unknown post-Newtonian coefficients (at 4PN, 5PN and 6PN orders) in the binding energy of the compact binary on a circular orbit. We shall also propose some gauge invariant definitions for the energy and angular momentum of a point particle in a binary system.

This paper is organized as follows. In Sec.~\ref{sec:PN} we complete the PN calculations performed in Papers I and II by taking into account new logarithmic contributions coming from memory-type hereditary terms in the metric, and show that including these new contributions yields the first law \eqref{first_law0}, which is thereby proved up to 3PN order plus the log-terms at 4PN and 5PN orders. In Sec.~\ref{sec:first_law} we show that this law is a particular case of the known general law of mechanics for systems of black holes and extended fluid bodies. Analogies with the single and binary black hole cases are drawn, and various discussions and alternative proofs are presented. Finally, in Sec.~\ref{sec:application} these results are applied (i) to provide gauge invariant candidates for the perturbed energy and angular momentum of a point particle in circular orbit about a Schwarzschild black hole, and (ii) to the numerical determination of high-order PN coefficients in the circular-orbit binding energy. We end up with a list of potential applications which are left for future work.

\section{Post-Newtonian derivation of the first law}
\label{sec:PN}

In Papers I and II, motivated by high-accuracy comparison between the post-Newtonian approximation and self-force (SF) perturbative results~\cite{De.08}, we derived the redshift observables $z_1$ and $z_2$ of a circular-orbit binary up to order 3PN, and included the specific contributions of logarithms at 4PN and 5PN orders (see also Ref.~\cite{Da.10} for the 4PN logarithm). In the limit of small mass ratio, we showed a very good agreement with numerical SF computations based on first-order black hole perturbation theory.

\subsection{Logarithmic contributions coming from memory-type hereditary terms}
\label{sec:log_contrib}

We have found that the previous calculation of the 4PN and 5PN logarithmic terms had overlooked a subtle point related to the assumption of helical symmetry: in the Appendix of Paper II, it was shown that a class of logarithmic terms contributing to the metric appeared in the form of an infinitesimal gauge transformation. These terms were left out of the calculation, because they could not contribute to the coordinate invariant relations computed for quasi-circular orbits. That conclusion is correct in the case of the physical problem where both conservative and dissipative effects are included, such that the binary decays and coalesces within a finite amount of time (from the point of view of a distant inertial observer). However, once the helical symmetry is imposed, the binary must be seen as having orbited for an infinite amount of time, and the infinitesimal gauge transformation used in the Appendix of Paper II turns out to become meaningless. In the present Section we correct for this effect and consider only a class of gauge transformations allowed by the HKV. We shall show that this adds extra logarithmic contributions at 4PN and 5PN orders, but that these do not affect the self-force regime investigated in Papers I and II.

The logarithms at 4PN and 5PN orders are produced by so-called ``hereditary'' terms~\cite{BlDa.92} when the helical symmetry is imposed. They can be computed by using the ``instantaneous'' propagator given by
\beq\label{propa}
\mathcal{I}^{-1} \equiv \mathop{\mathrm{FP}}_{B=0}\,\sum_{k=0}^{+\infty}\left(\frac{\partial}{c\partial t}\right)^{2k}\Delta^{-k-1}\left(\frac{r}{\lambda}\right)^B \,.
\eeq
This propagator depends explicitly on the orbital frequency $\Omega$ imposed by the helical symmetry, through the length scale $\lambda=2\pi c/\Omega$. Indeed a certain Finite Part (FP) operation at $B=0$ is required (see \textit{e.g.} Papers I and II), which involves the regulator $(r/\lambda)^B$ and yields when $B\to 0$ some logarithms of the type $\ln(r/\lambda)$. At quadratic non-linear order these logarithms are contained into the following piece of the ``gothic'' metric:\footnote{We pose $h^{\alpha\beta}\equiv\sqrt{-g}\,g^{\alpha\beta}-\eta^{\alpha\beta}$, where $g^{\alpha\beta}$ denotes the inverse of the usual covariant metric $g_{\alpha\beta}$, of determinant $g = \text{det}(g_{\alpha\beta})$, and $\eta^{\alpha\beta}$ is the Minkowski metric. We use harmonic coordinates throughout, \textit{i.e.} $\partial_\beta h^{\alpha\beta}=0$.}
\beq\label{deltah2}
\delta h^{\alpha\beta}_2 = \mathcal{I}^{-1}\left[\frac{1}{r^2}\left(\frac{4 M}{c^4}\,\mathop{z}^{(2)}\!{}^{\alpha\beta} + \frac{k^\alpha k^\beta}{c^2}\sigma \right)\right] . 
\eeq
The first term in the right-hand-side (RHS) corresponds to gravitational-wave tails (back-scattering of linear waves onto the background curvature associated with the monopole $M$ of the source), while the second term is due to the non-linear memory effect (radiation of waves by the stress-energy distribution of linear waves). The tensor $z^{\alpha\beta}(\bm{n},u)$ in Eq.~\eqref{deltah2} represents the coefficient of the dominant $1/r$ term of the (non-static part of the) linearized metric $h_1^{\alpha\beta}$ at future null infinity, \textit{i.e.} when $r\to+\infty$ with the retarded time $u\equiv t-r/c$ kept fixed. It is in the form of a sum of multipoles parametrized by mass-type moments $M_L(u)$ and current-type moments $S_L(u)$:
\begin{subequations}\label{zmunu}
\begin{align}
z^{00} &= - 4 \sum_{\ell\geqslant 2}\frac{n_L}{c^{\ell+2}\ell!}M_L^{(\ell)}(u)\,,\\
z^{0i} &= - 4 \sum_{\ell\geqslant 2}\left[\frac{n_{L-1}}{c^{\ell+2}\ell!}M_{iL-1}^{(\ell)}(u) - \frac{\ell}{c^{\ell+3}(\ell+1)!}\varepsilon_{iab}\,n_{aL-1}S_{bL-1}^{(\ell)}(u)\right] ,\\
z^{ij} &= - 4 \sum_{\ell\geqslant 2}\left[\frac{n_{L-2}}{c^{\ell+2}\ell!}M_{ijL-2}^{(\ell)}(u) -\frac{2\ell}{c^{\ell+3}(\ell+1)!}n_{aL-2}\,\varepsilon_{ab(i}S_{j)bL-2}^{(\ell)}(u)\right] .
\end{align}\end{subequations}
Our notation for multi-indices such as $L=i_1\cdots i_\ell$ (and so on) is the same as in Paper II. The superscript $(\ell)$ refers to time derivatives. In the second term of the RHS of Eq.~\eqref{deltah2}, $k^\alpha=(1,\bm{n})$ is the outgoing Minkowskian null vector and $\sigma(\bm{n},u)$, which is essentially the energy carried away by the waves in the direction $\bm{n}=\bm{x}/r$, is given by
\beq\label{sigma}
\sigma=\frac{1}{2}\mathop{z}^{(1)}\!{}^{\alpha\beta}\mathop{z}^{(1)}\!{}_{\alpha\beta}-\frac{1}{4}\mathop{z}^{(1)}\!{}^{\alpha}_{\alpha}\mathop{z}^{(1)}\!{}^{\beta}_{\beta}\,.
\eeq

The whole computation of the log-terms in Paper II was based on the first contribution (tail) in Eq.~\eqref{deltah2}. The second contribution (memory) was discarded because the log-terms therein are formally in the form of an infinitesimal gauge transformation. However, the argument overlooks the fact that the needed gauge transformation involves some \textit{anti-derivatives} of the quantity $\sigma$ defined by \eqref{sigma}; \textit{cf.} the Appendix of Paper II. Since the average of $\sigma$ over all directions represents the energy flux in the waves [see \textit{e.g.} Eq.~\eqref{energyflux} below], this means that the gauge transformation is actually not admissible in the presence of the helical symmetry, because the integral of the flux, or total energy emitted, is infinite in that case.

In the present paper we redo the analysis of the memory term in Eq.~\eqref{deltah2}, \textit{i.e.} 
\beq\label{deltahmem}
\bigl[\delta h^{\alpha\beta}_2\bigr]_\text{memory} = \mathcal{I}^{-1}\biggl[\frac{k^\alpha k^\beta}{r^2 c^2}\sigma(\bm{n},u)\biggr] \,,
\eeq
and show that it does contribute (together with its iteration at cubic non-linear order) to some new log-terms at 4PN and 5PN orders with respect to the results of Paper II. To this end we decompose $\sigma(\bm{n},u)$ into symmetric and trace-free (STF) spherical harmonics:
\beq\label{STF}
\sigma(\bm{n},u)=\sum_{\ell=0}^{+\infty} n_L\,\hat{\sigma}_L(u)\,,
\eeq
where $\hat{\sigma}_L(u)$ denotes the STF coefficient of order $\ell$ and $n_L=n_{i_1}\cdots n_{i_\ell}$. In the Appendix of Paper II, it is shown that the log-terms issued from Eq.~\eqref{deltahmem} are given by\footnote{We no longer mention the ``memory'' origin of this term. For the rest of this Section the notation $\delta q$ will refer to a modification in the quantity $q$ with respect to Paper II, \textit{i.e.} which is to be \textit{added} to the corresponding results of Paper II.}
\beq\label{memlog}
\delta h_2^{\alpha\beta} = \ln\left(\frac{r}{\lambda}\right)\sum_{\ell=0}^{+\infty} (-c)^{\ell+1} \partial^\alpha\partial^\beta\partial_{L}\bigl\{\hat{\sigma}^{(-\ell-3)}_{L}\bigr\}\,,
\eeq
where $\partial_L=\partial_{i_1}\cdots\partial_{i_\ell}$ and the superscript $(-\ell-3)$ refers to time \textit{anti}-derivatives. We employ the following convenient notation for a monopolar anti-symmetric wave built from an arbitrary function $F(u)$~\cite{Bl.93}:
\beq\label{notation}
\bigl\{F\bigl\} \equiv \frac{F(t-r/c) - F(t+r/c)}{2 r}\,.
\eeq
It was then shown in Paper II that the log-terms \eqref{memlog} can \textit{formally} be rewritten in the form of an infinitesimal gauge transformation, \textit{i.e.} $\delta h_2^{\alpha\beta} = 2 \partial^{(\alpha}\xi^{\beta)} - \eta^{\alpha\beta}\partial_\gamma\xi^\gamma$, where the gauge vector is explicitly given by
\beq\label{gauge}
\xi^\alpha = \frac{1}{2}\ln\left(\frac{r}{\lambda}\right)\sum_{\ell=0}^{+\infty} (-c)^{\ell+1} \partial^\alpha\partial_{L}\bigl\{\hat{\sigma}^{(-\ell-3)}_{L}\bigr\}\,.
\eeq

\subsubsection{``Maximal allowed'' gauge tranformation}

However it is crucial to control the occurence of time anti-derivatives in this gauge vector. Recall from Paper II that we are looking for so-called \textit{near-zone} logarithms, present in the near-zone expansion when $r\to 0$ of the metric. So we must here discard time anti-derivatives which appear in the near-zone expansion of the gauge vector \eqref{gauge}. The near-zone expansion when $r\to 0$ of the antisymmetric wave \eqref{notation} is regular, and given by
\beq\label{NZexp}
\bigl\{F\bigl\} = - \frac{1}{c} \sum_{k=0}^{+\infty} \frac{1}{(2k+1)!}\left(\frac{r}{c}\right)^{2k}F^{(2k+1)}(t)\,.
\eeq
Inspection of Eq.~\eqref{gauge} with the help of the expansion formula \eqref{NZexp} then shows that the \textit{monopole} $\ell=0$ and \textit{dipole} $\ell=1$ contributions in \eqref{gauge} involve time anti-derivatives. These are clearly incompatible with the imposition of the helical Killing symmetry, since for instance the anti-derivative of the energy flux is the total radiated energy, which is infinite. We shall therefore have to define a gauge transformation $\eta^\alpha$ differing from Eq.~\eqref{gauge}, and whose near-zone expansion is free of time anti-derivatives. There is a large number of possible choices for such a gauge vector. Here we choose to define $\eta^\alpha$ by removing from $\xi^\alpha$ a minimal number of terms containing the putative anti-derivatives, in such a way that the new gauge is still harmonic. This corresponds to a ``maximal allowed'' gauge transformation in the context of helical symmetric spacetimes. Our definition of $\eta^\alpha$ reads
\begin{subequations}\label{eta}
\begin{align}
\eta^0 &= \xi^0 - \frac{1}{2}\bigl\{\hat{\sigma}^{(-2)}\bigr\}\ln\left(\frac{r}{\lambda}\right) ,\\
\eta^i &= \xi^i - \frac{1}{6}\bigl\{\hat{\sigma}^{(-2)}_i\bigr\}\ln\left(\frac{r}{\lambda}\right) ,
\end{align}\end{subequations}
where $\hat{\sigma}$ and $\hat{\sigma}_i$ denote the monopolar and dipolar coefficients in the STF multipole expansion \eqref{STF}, given by (with $\ud\Omega$ the solid angle in the direction $\bm{n}$)
\begin{subequations}\label{hatsigma}
\begin{align}
\hat{\sigma}(u) &= \int\frac{\ud \Omega}{4\pi}\,\sigma(\bm{n},u) \,,\\
\hat{\sigma}_i(u) &= 3\int\frac{\ud \Omega}{4\pi}\,n_i\,\sigma(\bm{n},u)\,.
\end{align}\end{subequations}
An easy computation using the definition \eqref{sigma} together with the explicit expression \eqref{zmunu} of $z^{\alpha\beta}$ shows that $\hat{\sigma}(u)$, which is the average of $\sigma(\bm{n},u)$, starts to contribute at order $1/c^8$, while the dipole part $\hat{\sigma}_i(u)$ starts at order $1/c^9$. These will correspond to 4PN and 5PN terms in the metric, respectively. Here, to be consistent with 5PN, we have to obtain also the next-to-leading correction $\mathcal{O}(1/c^{10})$ in $\hat{\sigma}$, while the leading term is sufficient for $\hat{\sigma}_i$. We find
\begin{subequations}\begin{align}
\hat{\sigma} &= \frac{4}{5c^8}M^{(3)}_{ij}M^{(3)}_{ij}+\frac{1}{c^{10}}\left[\frac{4}{189}M^{(4)}_{ijk}M^{(4)}_{ijk}+\frac{64}{45}S^{(3)}_{ij}S^{(3)}_{ij}\right] + \mathcal{O}\biggl(\frac{1}{c^{12}}\biggr) \, , \label{energyflux}\\
\hat{\sigma}_i &= \frac{1}{c^{9}}\left[\frac{8}{21}M^{(3)}_{jk}M^{(4)}_{ijk}+ \frac{64}{15}\varepsilon_{ijk}M^{(3)}_{jl}S^{(3)}_{kl}\right] + \mathcal{O}\biggl(\frac{1}{c^{11}}\biggr) \, .
\end{align}\end{subequations}
We recognize the fluxes of energy and linear momentum carried away by the gravitational waves at future null infinity, so that the usual balance equations for the losses of energy $E$ and linear momentum $P_i$ read, with such notations,
\begin{subequations}\label{balance}
\begin{align}
\frac{\ud E}{\ud t} &= - \frac{G c^3}{4} \,\hat{\sigma}\,, \label{balanceE}\\
\frac{\ud P_i}{\ud t} &= - \frac{G c^2}{12} \,\hat{\sigma}_i\,.
\end{align}\end{subequations}
Since the energy flux is accurate up to next-to-leading order with respect to the quadrupole approximation, the quadrupole moment $M_{ij}$ in Eq.~\eqref{energyflux} has to properly include the 1PN relativistic corrections.

Performing the maximal allowed gauge transformation with gauge vector $\eta^\alpha$ defined by Eq.~\eqref{eta}, we obtain the following extra logarithmic contributions in the quadratic metric perturbation (with respect to Eqs.~(3.3) in Paper II):
\begin{subequations}\label{dh2munu}
\begin{align}
\delta h_2^{00} &= \biggl[-\frac{1}{2c}\bigl\{\hat{\sigma}^{(-1)}\bigr\} + \frac{1}{6}\partial_i\bigl\{\hat{\sigma}^{(-2)}_i\bigr\}\biggr]\ln\left(\frac{r}{\lambda}\right),\\
\delta h_2^{0i} &= \biggl[\frac{1}{2}\partial_i\bigl\{\hat{\sigma}^{(-2)}\bigr\} - \frac{1}{6c}\bigl\{\hat{\sigma}^{(-1)}_i\bigr\}\biggr]\ln\left(\frac{r}{\lambda}\right),\\
\delta h_2^{ij} &= \biggl[\frac{1}{3}\partial_{(i}\bigl\{\hat{\sigma}^{(-2)}_{j)}\bigr\} - \frac{1}{2c}\delta_{ij}\bigl\{\hat{\sigma}^{(-1)}\bigr\} - \frac{1}{6}\delta_{ij}\partial_k\bigl\{\hat{\sigma}^{(-2)}_k\bigr\}\biggr]\ln\left(\frac{r}{\lambda}\right).
\end{align}\end{subequations}
It can be checked that $\partial_\beta \delta h_2^{\alpha\beta}=0$ (modulo some non-logarithmic terms); hence the metric is still harmonic. The formula \eqref{NZexp} then gives the 4PN and 5PN terms in the metric perturbation as
\begin{subequations}\label{dh2munuexp}
\begin{align}
\delta h_2^{00} &= \biggl[\frac{1}{2c^2}\hat{\sigma} + \frac{r^2}{12c^4}\hat{\sigma}^{(2)} - \frac{x^k}{18c^3}\hat{\sigma}_k^{(1)}\biggr]\ln\left(\frac{r}{\lambda}\right) + \mathcal{O}\biggl(\frac{1}{c^{14}}\biggr)\,,\\
\delta h_2^{0i} &= \biggl[\frac{1}{6c^2}\hat{\sigma}_i - \frac{x^i}{6c^3}\hat{\sigma}^{(1)}\biggr]\ln\left(\frac{r}{\lambda}\right) + \mathcal{O}\biggl(\frac{1}{c^{13}}\biggr)\,,\\
\delta h_2^{ij} &= \biggl[\frac{1}{2c^2}\delta_{ij}\hat{\sigma} + \frac{r^2}{12c^4}\delta_{ij}\hat{\sigma}^{(2)} - \frac{1}{9c^3}x^{(i}\hat{\sigma}_{j)}^{(1)} + \frac{1}{18c^3}\delta_{ij}x^{k}\hat{\sigma}_{k}^{(1)}\biggr]\ln\left(\frac{r}{\lambda}\right)+ \mathcal{O}\biggl(\frac{1}{c^{14}}\biggr)\,.
\end{align}\end{subequations}
As usual this quadratic 4PN $+$ 5PN contribution will generate at cubic order an extra 5PN contribution, which is readily computed using techniques similar to those in Paper II. The required equation to be integrated is
\beq\label{eqh3}
\Delta\left[\delta h_3^{00} + \delta h_3^{ii}\right] = \frac{4}{c^2} \delta h_2^{ij}\partial_{ij}U + \frac{8}{c^2} \partial_i\delta h_2^{00}\partial_{i}U + \mathcal{O}\biggl(\frac{1}{c^{14}}\biggr)\,,
\eeq
where $U$ is the Newtonian potential obeying $\Delta U = -4\pi G\rho_*$.\footnote{Here $\rho_*=m_1\delta(\bm{x}-\bm{y}_1)+m_2\delta(\bm{x}-\bm{y}_2)$ is the Newtonian coordinate mass density of the particles.} At 4PN order both $\delta h_2^{00}$ and $\delta h_2^{ii}$ are proportional to $\hat{\sigma}$, and thus depend only on time (modulo some non log-terms). Using this fact we can readily integrate Eq.~\eqref{eqh3} and obtain the following cubic contribution to the log-terms at 5PN order:\footnote{We also obtain a contribution of far-zone logarithms given by
$$\left(\delta h_3^{00} + \delta h_3^{ii}\right)_\text{FZ} = \frac{2}{c^4}\hat{\sigma}\ln\left(\frac{r}{\lambda}\right)\sum_{\ell\geqslant 0}\frac{(-)^\ell}{\ell!(2\ell+1)}M_L\partial_L\left(\frac{1}{r}\right).
$$
However, according to the arguments in Paper II, we can ignore the contribution of far-zone logarithms.}
\beq
\delta h_3^{00} + \delta h_3^{ii} = \frac{2}{c^4} \,\hat{\sigma} \,U \ln\left(\frac{r}{\lambda}\right) + \mathcal{O}\biggl(\frac{1}{c^{14}}\biggr)\,.
\eeq
Finally, gathering the results we end up with the supplementary contributions (\textit{i.e.}, with respect to Eqs.~(3.9) and (3.12) in Paper II) of the logarithms at 4PN and 5PN orders in the covariant metric as
\begin{subequations}\label{gmunuharm}
\begin{align}
\delta g_{00} &= \biggl[-\frac{G^2}{c^2}\hat{\sigma} + \frac{2G^2}{c^4}U\,\hat{\sigma} - \frac{G^2}{6c^4}r^2\,\hat{\sigma}^{(2)} \biggr]\ln\left(\frac{r}{\lambda}\right) + \mathcal{O}\biggl(\frac{1}{c^{14}}\biggr)\,,\\
\delta g_{0i} &= \biggl[\frac{G^2}{6c^2}\hat{\sigma}_i - \frac{G^2}{6c^3}x^i\,\hat{\sigma}^{(1)}\biggr]\ln\left(\frac{r}{\lambda}\right) + \mathcal{O}\biggl(\frac{1}{c^{13}}\biggr)\,,\\
\delta g_{ij} &= \mathcal{O}\biggl(\frac{1}{c^{12}}\biggr)\,.
\end{align}\end{subequations}
This harmonic metric is rather simple and we have done all computations with it. 

\subsubsection{Cross-checking in a different gauge} 

As a check of the computations, we have also used an equivalent metric corresponding to a different (non-harmonic) coordinate system, and given by
\begin{subequations}\label{gmunuprime}
\begin{align}
\delta g'_{00} &= \biggl[-\frac{G^2}{c^2}\hat{\sigma} + \frac{2G^2}{c^4}U\,\hat{\sigma}\biggr]\ln\left(\frac{r}{\lambda}\right) + \Delta^{-1}\biggl[\frac{G^2}{c^4}\hat{\sigma}\,\Delta U \ln\left(\frac{r}{\lambda}\right)\biggr] + \mathcal{O}\biggl(\frac{1}{c^{14}}\biggr)\,,\label{g00prime}\\
\delta g'_{0i} &= \biggl[\frac{G^2}{6c^2}\hat{\sigma}_i - \frac{G^2}{2c^3}x^i\,\hat{\sigma}^{(1)}\biggr]\ln\left(\frac{r}{\lambda}\right) + \mathcal{O}\biggl(\frac{1}{c^{13}}\biggr)\,,\\
\delta g'_{ij} &= -\frac{G^2}{c^2}\delta_{ij}\hat{\sigma} \ln\left(\frac{r}{\lambda}\right) + \mathcal{O}\biggl(\frac{1}{c^{12}}\biggr)\,.
\end{align}\end{subequations}
This new metric involves a term which is not proportional to $\ln (r/\lambda)$ but instead contains the logarithm in its ``source'', originating in a 5PN modification of the source density $\rho_*$. This term, namely
\beq\label{lastterm}
\Delta^{-1}\biggl[\frac{G^2}{c^4}\hat{\sigma}\,\Delta U \ln\left(\frac{r}{\lambda}\right)\biggr] = \frac{G^2}{c^4}\hat{\sigma}\left[\frac{G m_1}{\vert\bm{x}-\bm{y}_1\vert}\ln\left(\frac{\vert\bm{y}_1\vert}{\lambda}\right)+\frac{G m_2}{\vert\bm{x}-\bm{y}_2\vert}\ln\left(\frac{\vert\bm{y}_2\vert}{\lambda}\right)\right] ,
\eeq
is crucial to take into account in the calculation with the alternative metric \eqref{gmunuprime}. When computing the equations of motion and the associated conserved quantities, we must apply the metric at the coordinate locations of the particles, so that $\ln (r/\lambda)$ becomes $\ln (\vert\bm{y}_1\vert/\lambda)$ or $\ln (\vert\bm{y}_2\vert/\lambda)$, which play the same role as the logarithms generated in Eq.~\eqref{lastterm}. In the center-of-mass frame, both types of logarithms become \textit{in fine} $\ln (r_{12}/\lambda)$ (modulo irrelevant constant terms), where $r_{12}=\vert\bm{y}_1-\bm{y}_2\vert$ is the coordinate separation. By contrast, in the computation with the original metric \eqref{gmunuharm}, there is no such term as \eqref{lastterm} corresponding to a logarithmic modification of the source. Rather, all terms in Eq.~\eqref{gmunuharm} are proportional to $\ln (r/\lambda)$. The two computations are however easily reconciled because the metrics \eqref{gmunuharm} and \eqref{gmunuprime} only differ by a coordinate transformation ${x'}^\alpha=x^\alpha+\epsilon^\alpha(x)$, namely
\begin{subequations}\label{diffgmunu}
\begin{align}
\delta g'_{00} &= \delta g_{00} -\frac{2}{c}\partial_t\epsilon_0 + \frac{2}{c^2}\left[\left(\epsilon^k-\epsilon_1^k\right)\frac{\partial U}{\partial y_1^k}+\left(\epsilon^k-\epsilon_2^k\right)\frac{\partial U}{\partial y_2^k}\right] + \mathcal{O}\biggl(\frac{1}{c^{14}}\biggr)\,,\label{diffg00}\\
\delta g'_{0i} &= \delta g_{0i} -\partial_i\epsilon_0-\frac{1}{c}\partial_t\epsilon_i + \mathcal{O}\biggl(\frac{1}{c^{13}}\biggr)\,,\\
\delta g'_{ij} &= \delta g_{ij} -\partial_i\epsilon_j - \partial_j\epsilon_i + \mathcal{O}\biggl(\frac{1}{c^{12}}\biggr)\,,
\end{align}\end{subequations}
where $\epsilon_A^k \equiv \epsilon^k(\bm{y}_A,t)$, with explicit expression for the gauge vector
\begin{subequations}\label{epsgauge}
\begin{align}
\epsilon^0 &= \frac{G^2}{12c^3}r^2\,\hat{\sigma}^{(1)} \,\ln\left(\frac{r}{\lambda}\right) + \mathcal{O}\biggl(\frac{1}{c^{13}}\biggr) \,,\\
\epsilon^i &= \frac{G^2}{2c^2}x^i\,\hat{\sigma} \,\ln\left(\frac{r}{\lambda}\right) + \mathcal{O}\biggl(\frac{1}{c^{12}}\biggr)\,.\end{align}\end{subequations}
The log-terms arising from the non-linear contribution in Eq.~\eqref{diffg00} are exactly given by \eqref{lastterm}, so that this term can be eliminated by the coordinate transformation \eqref{epsgauge}, and the two computations give the same gauge invariant results.

\subsection{Modification of the equations of motion and invariant quantities}

We then compute the modification of the acceleration of one of the particles induced by the change \eqref{gmunuharm} in the metric. For definiteness we use the harmonic coordinate system for this computation. Because the additional logarithmic contributions \eqref{gmunuharm} reduce to the mere function of time $\hat{\sigma}(t)$ at 4PN order, the modification of the acceleration occurs only at 5PN order. We find for the particle 1 (discarding uncontrolled non log-terms)
\beq\label{deltaa}
	\delta a_1^i = \biggl[- \hat{\sigma}\,\frac{G^3 m_2}{r_{12}^2 c^2} \, n_{12}^i + \frac{G^2}{2c^2} \hat{\sigma}^{(1)} \, v_1^i - \frac{G^2}{6c}\hat{\sigma}_i^{(1)} \biggr] \ln\left(\frac{r_{12}}{\lambda}\right) ,
\eeq
where $n^i_{12} = (y_1^i - y_2^i) / r_{12}$ and $v_1^i = \ud y_1^i/\ud t$ denotes the coordinate velocity. For simplicity we no longer write the neglected 6PN remainder. We now check that this modification of the acceleration is \textit{conservative}, in the sense that it corrects the conserved energy $E$, angular momentum $J_i$, linear momentum $P_i$, and center-of-mass position $G_i$ of the binary system. Indeed, a few calculations (see Paper II for more details) reveal that these corrections, which all occur at 5PN order, are given by
\begin{subequations}\label{deltacons}
\begin{align}
\delta E &= \biggl[- \frac{G^2}{2 c^2} \hat{\sigma}\,\Bigl(m_1 v_1^2+m_2 v_2^2\Bigr) + \frac{G^2}{6c} \hat{\sigma}_i\,M^{(1)}_i \biggr] \ln\left(\frac{r_{12}}{\lambda}\right) , \\
\delta J_i &= \biggl[- \frac{G^2}{2 c^2} \hat{\sigma}\,J_i + \frac{G^2}{6c} \varepsilon_{ijk} M_j \hat{\sigma}_k \biggr] \ln\left(\frac{r_{12}}{\lambda}\right) ,\\
\delta P_i &= \biggl[- \frac{G^2}{2 c^2} \hat{\sigma}\,M^{(1)}_i + \frac{G^2}{6c} \hat{\sigma}_i\,M \biggr] \ln\left(\frac{r_{12}}{\lambda}\right) ,\\
\delta G_i &= \biggl[\frac{2G}{c^5} \,E \,M^{(1)}_i\biggr] \ln\left(\frac{r_{12}}{\lambda}\right) ,
\end{align}\end{subequations}
where $M$ is the monopole of the source (reducing to $m=m_1+m_2$ at this level of approximation), $M_i=m_1 y_1^i + m_2 y_2^i$ is the binary's mass dipole, and $E$ is the relativistic 2.5PN-accurate binding energy satisfying the balance equation \eqref{balanceE}. We now work in the center-of-mass frame defined by $G_i=0$, which gives in particular $M_i=M_i^{(1)}=0$; hence
\begin{subequations}\label{deltaconsCM}
\begin{align}
\delta E &= - \frac{G^2}{2 c^2} \hat{\sigma}\,\Bigl(m_1 v_1^2+m_2 v_2^2\Bigr) \ln\left(\frac{r_{12}}{\lambda}\right) ,\\
\delta J_i &= - \frac{G^2}{2 c^2} \hat{\sigma}\,J_i  \ln\left(\frac{r_{12}}{\lambda}\right) , \\
\delta P_i &=  \frac{G^2}{6c} \hat{\sigma}_i\,M \ln\left(\frac{r_{12}}{\lambda}\right) ,\\
\delta G_i &= 0 \,.\label{deltaGi}
\end{align}\end{subequations}

For circular orbits, the extra log-terms in the relative acceleration $a^i = a_1^i - a_2^i$ reduce to
\beq\label{deltaai}
\delta a^i = -\frac{128}{5}\frac{G^6m^6}{r^7c^{10}} \, \nu^2 \, n^i \ln\left(\frac{r}{\lambda}\right) ,
\eeq
where at this stage we denote $r\equiv r_{12}$ and $n^i\equiv n_{12}^i$,\footnote{Note the slight inconsistency in notation here: $r=r_{12}$ is the binary's separation in \eqref{deltaai}, while $r=\vert\bm{x}\vert$ represented the distance to the field point in \textit{e.g.} Eqs.~\eqref{gmunuharm}.} and we have introduced the symmetric mass ratio $\nu=m_1m_2/m^2$, with $m=m_1+m_2$ the total mass. The result \eqref{deltaai} yields in turn a correction in the invariant orbital frequency $\Omega$ as a function of the parameter $\gamma\equiv G m/(r c^2)$ (see Paper II for notation) given by
\beq\label{deltaOm2}
    \delta\Omega^2 = \frac{G m}{r^3} \left[ \frac{64}{5} \,\nu^2\,\gamma^5\,\ln\gamma \right] ,
\eeq
where we have used the fact that $\ln(r/\lambda)=\frac{1}{2}\ln\gamma$ (see Paper II). From Eq.~\eqref{deltaOm2} we obtain the correction in $\gamma$ as a function of the orbital frequency $\Omega$, or rather of the dimensionless invariant PN parameter $x \equiv (G \, m \, \Omega/c^3)^{2/3}$, as
\beq\label{deltagam}
    \delta\gamma = x \left[-\frac{64}{15} \,\nu^2\,x^5\,\ln x \right] .
\eeq
It is important to realize that the results \eqref{deltaai}, \eqref{deltaOm2} and \eqref{deltagam} are coordinate dependent. They are given here in the harmonic gauge defined by the metric \eqref{gmunuharm}. Finally, making use of these last results, we find the following 5PN logarithmic corrections in the energy and angular momentum for circular orbits in the center-of-mass frame:\footnote{Beware that the notation $\delta E$ and $\delta J$ here has not exactly the same meaning as in Eqs.~\eqref{deltacons}--\eqref{deltaconsCM}. Indeed, there are additional pieces coming from the reduction to circular orbits of the Newtonian parts of $E$ and $J$ using Eqs.~\eqref{deltaOm2}--\eqref{deltagam}. However, because we have proved in \eqref{deltaGi} that $\delta G_i=0$, there is no additional piece coming from the passage to the center-of-mass frame. See also the discussion in Paper II.}
\begin{subequations}\label{deltaEJ}
\begin{align}
\delta E &= - \frac{1}{2} \, m\,\nu \, c^2\,x \biggl[ - \frac{64}{15} \, \nu^2 x^5 \ln{x} \biggr] \,,\label{deltaE}\\ 
\delta J &= \frac{G \,m^2 \,\nu}{c\,x^{1/2}} \biggl[ \frac{32}{15} \, \nu^2 x^5 \ln{x} \biggr] \,. \label{deltaJ}
\end{align}
\end{subequations}
These last results are coordinate invariant, and we have checked that they come out the same with either metric \eqref{gmunuharm} or \eqref{gmunuprime}. The corrections \eqref{deltaEJ} are such that $\delta E = \Omega \, \delta J$, which is natural because they arise from terms connected to the gravitational-wave fluxes [remember Eqs.~\eqref{balance}]. These 5PN logarithmic contributions to the energy and angular momentum  have to be added to the 4PN and 5PN terms already found in Paper II. Note that these corrections are proportional to $\nu^2$ and therefore affect the results only starting at the \textit{post-self-force} level. 

Next, we compute the redshift observable, which is coordinate invariant for circular orbits under the assumption of helical symmetry~\cite{De.08}. In a coordinate system in which the HKV reads $K^\alpha\partial_\alpha=\partial_t+\Omega\,\partial_\varphi$, it is given by the time component of the four-velocity of the particle:
\beq\label{u1t}
u^t_1 = \biggl(- g_{\alpha\beta}(y_1)\frac{v_1^\alpha v_1^\beta}{c^2}\biggr)^{-1/2} \,,
\eeq
where $v_1^\alpha = (c, v_1^i)$. Inserting the modification of the metric \eqref{gmunuharm}, together with the standard 1PN metric for consistent reduction to circular orbits making use of the results \eqref{deltaOm2} and \eqref{deltagam}, we find the 4PN+5PN corrections
\beq\label{deltau1t}
\delta u_1^t = \left[ - \frac{32}{5} + \left( \frac{1886}{105} - \frac{608}{105} \Delta + \frac{1592}{105} \nu \right) x \right] \nu^2 \, x^5 \ln{x} \, .
\eeq
We introduced the notation $\Delta \equiv (m_2-m_1)/m = \sqrt{1-4\nu}$ for the relative mass difference,\footnote{We assume, without any loss of generality, that $m_1 \leqslant m_2$.} so that $\delta u_2^t$ is simply obtained by changing $\Delta$ into $-\Delta$. Again, the correction \eqref{deltau1t} occurs at the post-self-force level, and does not affect the high-accuracy comparison between the post-Newtonian and the perturbative self-force calculations reported in Papers I and II.

In the following it will be more convenient to work with the inverse of $u_1^t$, denoted
\beq\label{z1}
z_1 \equiv \frac{1}{u^t_1} = \biggl(- g_{\alpha\beta}(y_1)\frac{v_1^\alpha v_1^\beta}{c^2}\biggr)^{1/2}\,,
\eeq
and which we will still refer to as the redshift observable following Ref.~\cite{De.08}. In terms of this variable the correction with respect to paper II reads\footnote{We take into account the leading-order terms in the expansion as given by $u_1^t = 1 + \left(\frac{3}{4}+\frac{3}{4}\Delta-\frac{\nu}{2}\right) x + \mathcal{O}(x^2)$.}
\beq\label{deltaz1}
\delta z_1 = \left[ \frac{32}{5} + \left( -\frac{2894}{105} - \frac{80}{21} \Delta - \frac{184}{21} \nu \right) x \right] \nu^2 \, x^5 \ln{x} \, .
\eeq

\subsection{Post-Newtonian results for the conserved quantities and redshift observable}
\label{subsec:PN_calculations}

The expressions $E(\Omega)$ and $J(\Omega)$ of the PN binding energy and angular momentum for point-particle binaries on quasi-circular orbits have been computed up to 3PN order by different groups~\cite{Da.al.00,BlFa2.00,Da.al.01,BlFa.01,Bl.al.04,ItFu.03}. More recently, the 3PN expansion of the redshift observable $z_1(\Omega)$ was computed in Paper I, and compared to the numerical result for the gravitational SF in linear black hole perturbation theory.

The dominant logarithmic contributions, which arise at 4PN order~\cite{An.al.82,BlDa.88}, together with the next-to-leading order 5PN contributions, were computed in Paper II. These were assumed to only come from gravitational-wave tails, heuristically the scattering of gravitational radiation by the background curvature generated by the monopole of the source. Furthermore, we have now computed some additional 4PN and 5PN logarithmic terms which come from the non-linear memory effect, heuristically the gravitational radiation generated by the stress-energy distribution of linear waves. All those logarithmic contributions are appropriate to conservative helically symmetric PN spacetimes.

Combining the result (4.12) of Paper II with the new correction \eqref{deltaE} obtained above we obtain the 5PN-accurate expression for the binding energy: 
\begin{align}
	E &= - \frac{1}{2} \, m\,\nu \,c^2\, x \left\{ 1 + \left( - \frac{3}{4} - \frac{\nu}{12} \right) x + \left( - \frac{27}{8} + \frac{19}{8} \nu - \frac{\nu^2}{24} \right) x^2 \right. \nonumber \\ &\qquad\qquad\qquad\quad\, + \left( - \frac{675}{64} + \biggl[ \frac{34445}{576} - \frac{205}{96} \pi^2 \biggr] \nu - \frac{155}{96} \nu^2 - \frac{35}{5184} \nu^3 \right) x^3 \nonumber \\ &\qquad\qquad\qquad\quad\, +\left( - \frac{3969}{128} + \nu \, e_4(\nu) + \frac{448}{15} \nu \ln{x} \right) x^4 \nonumber \\ & \left. \qquad\qquad\qquad\quad\, + \left( - \frac{45927}{512} + \nu \, e_5(\nu) + \biggl[ - \frac{4988}{35} - \frac{656}{5} \nu \biggr] \nu \ln{x} \right) x^5 \right\} , \label{E_x} 
\end{align}
where we recall that $x=(G m \Omega/c^3)^{2/3}$ (for simplicity we do not indicate the neglected 6PN remainder). We introduced some unknown 4PN and 5PN coefficients $e_4(\nu)$ and $e_5(\nu)$, which however are known to be \textit{polynomials} in the symmetric mass ratio.\footnote{The latter point can be proved from the fact that the expression of $E$ (and similarly for $J$ and $z_1$) for general orbits, \textit{i.e.} before restriction to the center-of-mass frame and circular orbits, must be a polynomial in the two separate masses $m_1$ and $m_2$.} In Sec.~\ref{E_PN} we shall be able to obtain precise numerical estimates of $e_4(0)$ and $e_5(0)$, which encode information at leading order beyond the test-particle result (corrections linear in $\nu$).

The computation of the angular momentum proceeds in the same way. We first derive the 4PN and 5PN logarithmic terms due to gravitational-wave tails, similarly to Section IV of Paper II, and then add the correction terms given by Eq.~\eqref{deltaJ}. The result is
\begin{align}
    J &= \frac{G \,m^2 \,\nu}{c\,x^{1/2}} \left\{ 1 + \left( \frac{3}{2} + \frac{\nu}{6} \right) x + \left( \frac{27}{8} - \frac{19}{8} \nu + \frac{\nu^2}{24} \right) x^2 \right. \nonumber \\ &\qquad\qquad\, + \left( \frac{135}{16} + \biggl[ - \frac{6889}{144} + \frac{41}{24} \pi^2 \biggr] \nu + \frac{31}{24} \nu^2 + \frac{7}{1296} \nu^3 \right) x^3 \nonumber \\ &\qquad\qquad\, +\left( \frac{2835}{128} + \nu \, j_4(\nu) - \frac{64}{3} \nu \ln{x} \right) x^4 \nonumber \\ & \left. \qquad\qquad\, + \left( \frac{15309}{256} + \nu \, j_5(\nu) + \biggl[ \frac{9976}{105} + \frac{1312}{15} \nu \biggr] \nu \ln{x} \right) x^5 \right\} , \label{J_x} 
\end{align}
where $j_4(\nu)$ and $j_5(\nu)$ denote other unknown coefficients which are also polynomials in $\nu$. Next, combining the results (4.10) of Paper I and (5.2) of Paper II with the correction term \eqref{deltau1t} found above, we obtain the 5PN-accurate expression of the redshift observable $u_1^t$, or rather of its inverse $z_1 = 1/u_1^t$, as
\begin{align}
	z_1 &= 1 + \left( - \frac{3}{4} - \frac{3}{4} \Delta + \frac{\nu}{2} \right) x + \left( - \frac{9}{16} - \frac{9}{16} \Delta - \frac{\nu}{2} - \frac{1}{8} \Delta \, \nu + \frac{5}{24} \nu^2 \right) x^2 \nonumber \\ &\qquad\! + \left( - \frac{27}{32} - \frac{27}{32} \Delta - \frac{\nu}{2} + \frac{19}{16} \Delta \, \nu - \frac{39}{32} \nu^2 - \frac{1}{32} \Delta \, \nu^2 + \frac{\nu^3}{16} \right) x^3 \nonumber \\ &\qquad\! + \left( - \, \frac{405}{256} - \frac{405}{256} \Delta + \left[ \frac{38}{3} - \frac{41}{64} \pi^2 \right] \nu + \left[ \frac{6889}{384} - \frac{41}{64} \pi^2 \right] \Delta \, \nu \right. \nonumber \\ &\qquad\qquad\!\! \left. + \left[ - \frac{3863}{576} + \frac{41}{192} \pi^2 \right] \nu^2 - \frac{93}{128} \Delta \, \nu^2 + \frac{973}{864} \nu^3 - \frac{7}{1728} \Delta \, \nu^3 + \frac{91}{10368} \nu^4 \right) x^4 \nonumber \\ &\qquad\! + \left( - \, \frac{1701}{512} - \, \frac{1701}{512} \Delta + \nu \, \bigl[p_4(\nu) + \Delta\,q_4(\nu)\bigr] + \left[ \frac{32}{5} + \frac{32}{5} \Delta + \frac{32}{15} \nu \right] \nu \, \ln x \right) x^5 \nonumber \\ &\qquad\! + \left( - \, \frac{15309}{2048} - \, \frac{15309}{2048}\Delta + \nu \, \bigl[p_5(\nu) + \Delta\,q_5(\nu)\bigr] \right. \nonumber \\ &\qquad\qquad\!\! \left. + \left[ - \frac{2494}{105} - \frac{2494}{105} \Delta - \frac{5938}{105} \nu - \frac{164}{5} \Delta \nu + \frac{328}{15} \nu^2 \right] \nu \, \ln x \right) x^6 \, . \label{z1_x}
\end{align}
The redshift observable $z_2$ of particle $2$ can immediately be deduced from $z_1$ by setting $\Delta \to - \Delta$. Here $p_4(\nu)$, $q_4(\nu)$ and $p_5(\nu)$, $q_5(\nu)$ denote some still further \textit{a priori} unknown 4PN and 5PN polynomials in the symmetric mass ratio.

Note that in each of Eqs.~\eqref{E_x}--\eqref{z1_x} we have added to the usual PN results, valid for any mass ratio, the 4PN and 5PN contributions from the test-mass limit for one of the particles, known from exact calculations for test particles in the Schwarzschild geometry:
\begin{subequations}\label{schw}
\begin{align}
	E &= m\,\nu\,c^2\left(\frac{1-2x}{\sqrt{1-3x}} - 1\right) + \mathcal{O}(\nu^2)\,,\label{Eschw}\\
	J &= \frac{G \,m^2 \,\nu}{c\,\sqrt{x(1-3x)}} + \mathcal{O}(\nu^2)\,, \label{Jschw}\\[0.8em]
	z_1 &= \sqrt{1-3x} + \mathcal{O}(\nu) \,.\label{z1schw}
\end{align}
\end{subequations}

\subsection{Post-Newtonian first law of black hole binaries}
\label{sec:PNfirstlaw}

In this Section we shall show that the previous post-Newtonian results for the energy $E$, angular momentum $J$, and redshift observables $z_{1,2}$ of binary systems on circular orbits satisfy a very important property called the first law of binary black hole mechanics. Here, since we are working within the PN framework, the appropriate description of (non-spinning) black holes is by structureless point masses; hence we shall also refer to this result as the first law of binary \textit{point-particle} mechanics.

For this study we shall introduce the total relativistic (ADM) mass of the binary system:\footnote{From now on we use geometrized units $G = c = 1$.}
\beq\label{PNmass}
M = m + E \, , 
\eeq
where $m=m_1+m_2$ is the sum of the post-Newtonian individual masses, \textit{i.e.} those which enter as coefficients of Dirac delta-functions in the stress-energy tensor of point particles; \textit{cf.} Eq.~\eqref{Tmunu} below. The various concepts of mass we use will be further specified in Sec.~\ref{subsec:notions_mass}.

The ADM mass $M$, total angular momentum $J$, and redshifts $z_{1,2}$, all being given by Eqs.~\eqref{E_x}--\eqref{z1_x}, are functions of the three independant variables of the problem, namely the orbital frequency $\Omega$ that is imposed by the existence of the HKV, and the individual masses $m_{1,2}$ of the particles. We first find, with the above expressions, that the ADM quantities obey the usual relation commonly used in PN theory (see \textit{e.g.} Refs.~\cite{Da.al.00,Bl.02}):
\beq\label{partial_Omega}
\frac{\partial M}{\partial \Omega} = \Omega \, \frac{\partial J}{\partial \Omega}\,.
\eeq
This well-known relation is, for instance, extensively used in computations of the binary evolution based on a sequence of quasi-equilibrium configurations~\cite{Go.al.02,Gr.al.02}. For black hole binaries moving on quasi-circular orbits, it states that the gravitational-wave energy and angular momentum fluxes are proportional (with $\Omega$ being the coefficient of proportionality). Here we find that this relation is satisfied for all the terms explicitly computed up to 5PN order, including the non trivial 4PN and 5PN logarithmic contributions.

Largely unrecognized however are the relations which tell how the ADM quantities change when the individual masses $m_1$ and $m_2$ of the particles vary (keeping the orbital frequency $\Omega$ fixed). By direct partial differentiation of the expressions \eqref{E_x}--\eqref{J_x} with respect to $m_1$ and $m_2$, and comparison with \eqref{z1_x}, we find
\begin{subequations}\label{PDE}
	\begin{align}
		\frac{\partial M}{\partial m_1} - \Omega \frac{\partial J}{\partial m_1} &= z_1 \, , \label{partial_m1} \\
		\frac{\partial M}{\partial m_2} - \Omega \frac{\partial J}{\partial m_2} &= z_2 \,. \label{partial_m2}
	\end{align}
\end{subequations}
Again, these relations hold for all the terms we have computed, namely up to 3PN order and including the 4PN and 5PN log-terms. Taking further partial derivatives of Eqs.~\eqref{PDE} with respect to the masses yields the simple symmetric relation
\beq\label{z1z2}
	\frac{\partial z_1}{\partial m_2} = \frac{\partial z_2}{\partial m_1} \, ,
\eeq
which can be viewed as reflecting some ``equilibrium'' state of the binary system.

The three relations \eqref{partial_Omega}--\eqref{PDE} can be summarized in the first law of the binary black hole (or binary point-particle) mechanics
\beq\label{first_law}
	\delta M - \Omega \, \delta J = z_1 \, \delta m_1 + z_2 \, \delta m_2 \,.
\eeq
This first law provides the changes in the ADM mass and angular momentum in response to infinitesimal variations of the individual masses of the point particles, weighted by their redshift factors. As we shall prove in Sec.~\ref{subsec:proof} that this law should be correct at all PN orders, its verification provides a very powerful test of the intricate post-Newtonian calculations yielding Eqs.~\eqref{E_x}--\eqref{z1_x}. An interesting by-product of the first law \eqref{first_law} is the remarkably simple relation
\beq\label{first_integral}
	M - 2 \Omega J = m_1 z_1 + m_2 z_2 \, ,
\eeq
which can be seen as a ``first integral'' of the differential law \eqref{first_law}. It should be stressed that the existence of such a simple, linear, algebraic relation between the local quantities $z_{1,2}$ on one hand, and the global quantities $M$ and $J$ on the other hand, is not trivial.\footnote{Interestingly, a result equivalent to \eqref{first_integral} was implicitly derived, in the case of a specific point-particle model, at first order in post-Minkowskian gravity: \textit{cf.} Eqs.~(92) and (96) of Ref.~\cite{FrUr.06}.}

In order to prove Eq.~\eqref{first_integral}, we take a linear combination of the two equations \eqref{PDE}, make the change of variables $(\Omega,m_1,m_2) \to (\Omega,m,\nu)$, where we recall that $m = m_1 + m_2$ is the total mass and $\nu = m_1 m_2 / m^2$ the symmetric mass ratio, to get
\beq\label{m1z1m2z2}
	m \frac{\partial \mathcal{M}}{\partial m} = m_1 z_1 + m_2 z_2 \, ,
\eeq
where we introduced the convenient combination
\beq\label{calM}
	 \mathcal{M} \equiv M - \Omega J \, ,
\eeq
which can heuristically be viewed as the energy of the binary in a co-rotating frame. Next, we notice that the ratio $\mathcal{M}/m$ is dimensionless and symmetric by exchange $m_1 \longleftrightarrow m_2$ of the masses. It must therefore depend on the three independant variables $(\Omega,m_1,m_2)$ only through the symmetric mass ratio $\nu$ and the product $m \Omega$. This last observation implies that
\beq\label{mOmega}
	m \frac{\partial (\mathcal{M}/m)}{\partial m} = \Omega \frac{\partial (\mathcal{M}/m)}{\partial \Omega} \, .
\eeq
Combining \eqref{m1z1m2z2} with \eqref{mOmega}, and making use of Eq.~\eqref{partial_Omega}, we obtain the relation \eqref{first_integral}. Alternatively, we can use Euler's theorem for homogeneous functions: since, from Eq.~\eqref{first_law}, the ADM mass $M$ must be a homogeneous function of degree one in $J^{1/2}$, $m_1$ and $m_2$, we have\footnote{Here the partial derivatives are to be taken using the independent variables $(J, m_1, m_2)$ instead of the variables $(\Omega, m_1, m_2)$ as used in Eqs.~\eqref{partial_Omega}--\eqref{PDE}. In this respect notice that
$$z_1=\frac{\partial M}{\partial m_1}\bigg|_{J,m_2}=\frac{\partial M}{\partial m_1}\bigg|_{\Omega,m_2}-\Omega\frac{\partial J}{\partial m_1}\bigg|_{\Omega,m_2}\,,$$
together with a similar expression for $z_2$.}
\beq
	M(J,m_1,m_2) = J^{1/2} \frac{\partial M}{\partial J^{1/2}} + m_1 \frac{\partial M}{\partial m_1} + m_2 \frac{\partial M}{\partial m_2} \, ,
\eeq
which combined with the first law \eqref{first_law} immediately yields the result \eqref{first_integral}. In Sec.~\ref{sec:Komar} we provide a third derivation, based on the integration of the Einstein field equations.

Note that we have explicitly verified the partial differential equations \eqref{partial_Omega} and \eqref{PDE}, and therefore the first law \eqref{first_law}, only for those terms in the ADM quantities $M$, $J$ and the redshifts $z_{1,2}$ which are known, namely for all the terms up to 3PN order and for the log-terms occuring at 4PN and 5PN orders. Evidently we could not verify the law for the 4PN and 5PN non-logarithmic contributions (and all 6PN and higher order terms) which are left out in the PN calculation. However it is not difficult to find the relationships which must be satisfied by the polynomials of the symmetric mass ratio parametrizing the unknown terms in Eqs.~\eqref{E_x}--\eqref{z1_x}. We find that Eq.~\eqref{partial_Omega} is exactly satisfied at 5PN order if and only if the unknown functions $j_4(\nu)$, $j_5(\nu)$ in the angular momentum are given in terms of the unknowns $e_4(\nu)$, $e_5(\nu)$ in the energy by
\begin{subequations}\label{j45}
\begin{align}
j_4(\nu) &= - \frac{5}{7} e_4(\nu) +\frac{64}{35} \,,\\
j_5(\nu) &= - \frac{2}{3} e_5(\nu) -\frac{4988}{945} - \frac{656}{135} \nu \,.
\end{align}
\end{subequations}
Furthermore, we find that the first law is fully satisfied up to 5PN order if and only if the unknowns $p_4(\nu)$, $p_5(\nu)$ and $q_4(\nu)$, $q_5(\nu)$ in the redshift \eqref{z1_x} are also given in terms of the functions $e_4(\nu)$, $e_5(\nu)$ in the energy, and their $\nu$-derivatives $e'_4(\nu)$, $e'_5(\nu)$, by
\begin{subequations}\label{pq45}
\begin{align}
p_4(\nu) &= \frac{13}{14}\nu \,e_4(\nu) + \left(1-4\nu\right)q_4(\nu) - \frac{3969}{256} - \frac{128}{35} \nu \,,\\
p_5(\nu) &= \frac{5}{6}\nu \,e_5(\nu) + \left(1-4\nu\right) q_5(\nu) - \frac{45927}{1024} + \frac{9976}{945} \nu + \frac{1312}{135} \nu^2 \,,\\
q_4(\nu) &= \frac{3}{28}\nu \,e'_4(\nu) + \frac{3}{14} e_4(\nu) - \frac{64}{35} \,,\\
q_5(\nu) &= \frac{1}{12}\nu \,e'_5(\nu) + \frac{1}{6} e_5(\nu) + \frac{4988}{945} + \frac{328}{45} \nu \,.
\end{align}
\end{subequations}
Thus, all unknown non logarithmic terms at 4PN and 5PN orders are parametrized by only two polynomials of the symmetric mass ratio, which can conveniently be chosen to be $e_4(\nu)$, $e_5(\nu)$ introduced in the binding energy \eqref{E_x}. Given the high physical significance of the first law, as well as the general poof of it given in Sec.~\ref{sec:first_law} (which does not rely on a PN expansion), we shall take for granted that it is valid for all PN terms, including the unknown ones, so the relationships \eqref{j45}--\eqref{pq45} must be exactly satisfied.  We will make further use of these relations in Sec.~\ref{sec:application}.

In summary, we have proved {by a high-order post-Newtonian calculation in this Section (building on the results of Papers I and II),} that the first law of black hole mechanics \eqref{first_law} holds in the particular case of a binary system moving on an exact circular orbit. We shall further demonstrate, in Sec.~\ref{sec:first_law}, that the law \eqref{first_law} is actually a particular case, valid when one assumes the existence of a HKV covering the entire spacetime, of the generalized law of black hole mechanics obtained by Friedman, Ury{\=u}, and Shibata~\cite{Fr.al.02}. Before doing so, we want to clarify the concepts of ADM and Bondi masses as we use them in post-Newtonian theory, \textit{i.e.} in relation to the relativistic mass $M$ introduced in Eq.~\eqref{PNmass}. 

\subsection{Bondi mass versus ADM mass in post-Newtonian theory}
\label{subsec:notions_mass}

The structure of the gravitational radiation field generated by isolated systems in general relativity was elucidated during the early sixties by Bondi, Sachs, Penrose and coworkers, who analyzed its asymptotic structure at future null infinity. Of particular importance is the Bondi-Sachs mass-loss formula~\cite{Bo.al.62,Sa.62}, which relates the rate of change of the Bondi mass $M_\text{B}(U)$ at a retarded instant of time $U$,\footnote{Here $U$ denotes a null retarded coordinate, which differs from the shorthand $u=t-r/c$ used in Sec.~\ref{sec:log_contrib}, where $\{t,r\}$ are harmonic coordinates.} to the gravitational-wave energy flux $\mathcal{F}$, through
\beq\label{mass_loss}
	\frac{\ud M_\text{B}}{\ud U} = - \mathcal{F}(U) \, ,
\eeq
where $\mathcal{F} = \oint_{\scri^+} \vert \mathcal{N} \vert^2 \, \ud \Omega$ is computed as a surface integral (at future null infinity) of the News function $\mathcal{N}$. Another important result was obtained later by Ashtekar and Magnon-Ashtekar~\cite{AsMa.79}, who showed that (for isolated gravitating systems) the difference between the ADM mass $M_\text{ADM}$ and the Bondi mass $M_\text{B}(U)$ is equal to the mass carried away by the gravitational radiation emitted between the infinite past and the given retarded instant:
\beq\label{Madm_Mb}
	M_\text{ADM} = M_\text{B}(U) + \int_{-\infty}^U \!\! \mathcal{F}(U') \, \ud U' \, .
\eeq
Since the ADM mass is constant, $\ud M_\text{ADM} / \ud U = 0$, the mass-loss formula \eqref{mass_loss} immediately follows from Eq.~\eqref{Madm_Mb}. More generally, similar results hold for the ADM four-momentum $P^\alpha = (M_\text{ADM},P^i)$ and the Bondi four-momentum $P_\text{B}^\alpha = (M_\text{B},P_\text{B}^i)$.

In addition to the ADM mass and the Bondi mass, many alternative notions of mass have been introduced in general relativity (see \textit{e.g.}~\cite{JaGo.11} for a recent review). However, while the definitions of $M_\text{ADM}$ and $M_\text{B}$ only rely on some \textit{universal} properties of spatial infinity and future null infinity, respectively, most of these alternative notions of mass require the introduction of one or several additional structure(s) on top of the spacetime metric $g_{\alpha\beta}$; the Komar mass $M_\text{K}$, defined in terms of the timelike Killing vector $t^\alpha$ of a stationary spacetime, being one specific example (\textit{cf.} Sec.~\ref{sec:Komar}).

In post-Newtonian theory, a background Minkowski metric $\eta_{\alpha\beta}$ is introduced in addition to the usual spacetime metric $g_{\alpha\beta}$. While this, in effect, breaks the manifest general covariance of the exact theory, the resulting field equations remain covariant under the Poincar\'e group of special relativity. In particular, the PN equations of motion of compact binary systems (in harmonic coordinates which preserve the Poincar\'e symmetry) are invariant under time translations, spatial translations, spatial rotations, and Lorentz boosts. Noether's theorem then implies the existence of ten conserved quantities associated with these continuous symmetries: the post-Newtonian binding energy $E$, linear momentum $P^i$, angular momentum $J^i$, and the vector $K^i$ such that the center-of-mass position reads $G^i = K^i + P^i t$.

The problem of constructing gravitational-wave templates for circularized inspiralling compact binary systems in the PN approximation is usually divided into two sub-problems: (i) the computation of the center-of-mass binding energy $E$, using the \textit{conservative} part of the dynamics of the source, and (ii) the calculation of the gravitational-wave energy flux $\mathcal{F}$, associated with the \textit{dissipative} part of the dynamics, from the wave-zone gravitational field. One then postulates the energy balance relation
\beq\label{energy_balance}
	\frac{\ud E}{\ud T} = - \mathcal{F}(T) \, ,
\eeq
which states that the binding energy decreases at a rate determined by the flux of energy carried away by the gravitational radiation. The coordinate time $T$ coincides with the proper time of an inertial observer far away from the source, where the geometry is essentially flat. In the case of compact binary systems whose components are modelled as point masses, both $E$ and $\mathcal{F}$ have been computed up to 3.5PN order included (see~\cite{Bl.06} for a review). One could in principle verify the energy balance relation \eqref{energy_balance} explicitly from the knowledge of all conservative and dissipative contributions to the dynamics of the binary system up to 3.5PN beyond the leading order radiation-reaction force, which is at leading 2.5PN order; this would correspond to 6PN order beyond the Newtonian motion. Since it is, in practice, too challenging to compute the PN equations of motion to such high orders, the relation \eqref{energy_balance} has to be assumed, rather than derived. Note however that Eq.~\eqref{energy_balance} has been proved to hold up to the relative 1.5PN order, for a generic matter source~\cite{BlDa.88,Bl.93,Bl.97}.

\begin{figure}[t!]
	\includegraphics[scale=0.3]{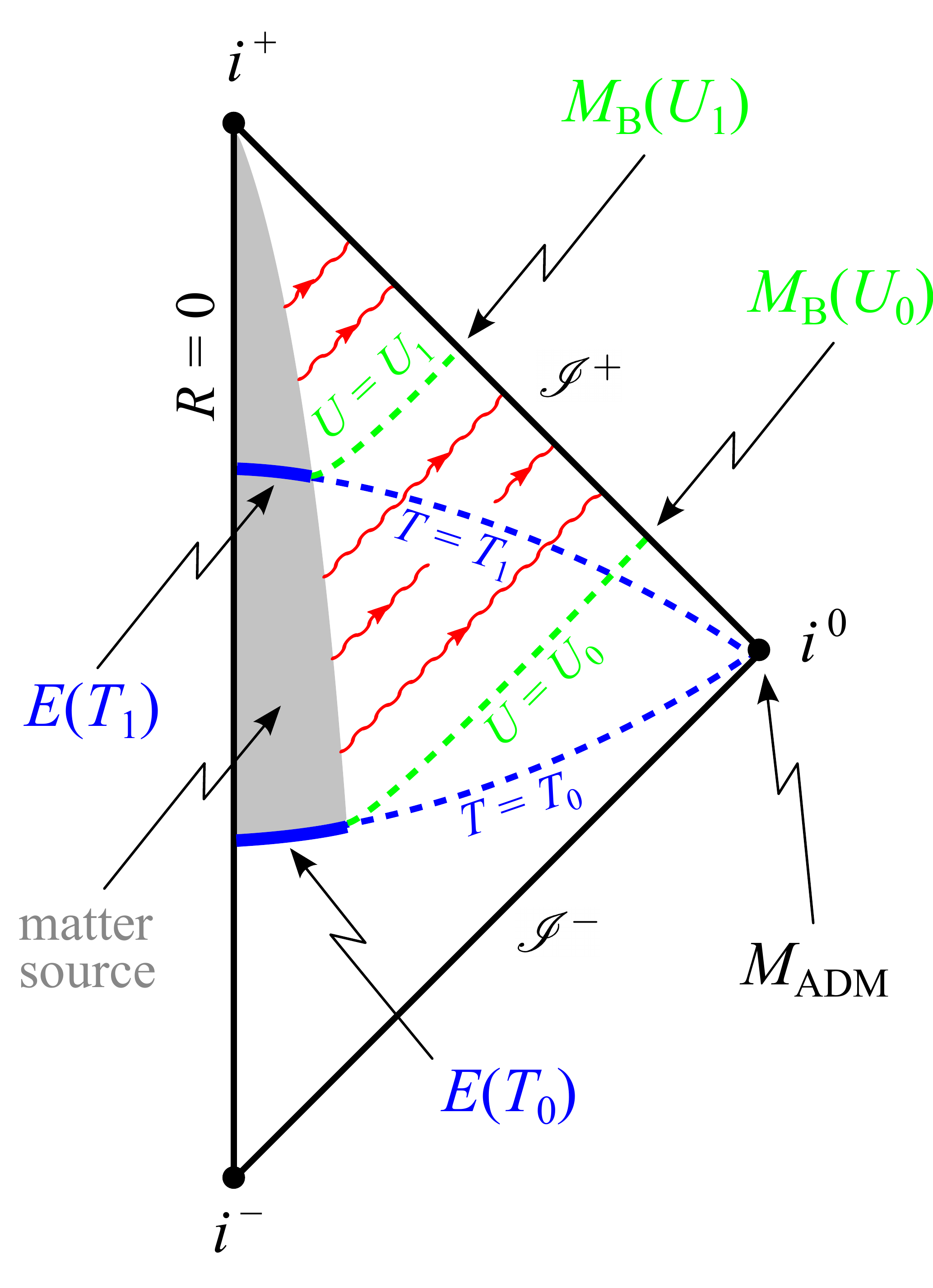}
	\caption{\footnotesize A gravitationally bound isolated matter source is formed at $T = T_0$, and starts emitting gravitational radiation. The ADM mass $M_\text{ADM}$, as computed on that time slice, coincides with the binding energy $M(T_0) = m + E(T_0)$, which is itself equal to the Bondi mass $M_\text{B}(U_0)$, as computed on the asymptotically null hypersurface $U = U_0$. At a later time $T = T_1$, the binding energy $M(T_1) = M_\text{B}(U_1)$ has decreased. The difference $M_\text{ADM} - M_\text{B}(U_1)$ with respect to the constant ADM mass is equal to the energy taken away from the source by the gravitational waves emitted between $T = T_0$ and $T = T_1$, or equivalently between $U = U_0$ and $U = U_1$.}
	\label{fig:cons_diss}
\end{figure}

The similarity between the PN energy balance equation \eqref{energy_balance} and the mass-loss formula \eqref{mass_loss} is of course striking. However, the mathematical objects entering these two relations are conceptually different: for asymptotically flat spacetimes, the Bondi mass $M_\textit{B}$ is defined in the exact theory as a surface integral at future null infinity, while the binding energy $E$ is only defined in the post-Newtonian approximation to general relativity, and its computation involves the near-zone PN metric. Although it has not been proved rigorously that the two notions of mass coincide, the identification of the Bondi mass with the PN binding energy (or, rather, with $M=m+E$ which includes the rest mass $m = m_1 + m_2$) seems intuitively sound and natural. Henceforth, we shall thus postulate that if $U = T - R$ is the asymptotically null outgoing coordinate associated with an asymptotically radiative (or Bondi-type) coordinate system $\{T,R\}$, then there exists a spacelike hypersurface $T = \text{const}$ such that
\beq\label{Mb_E}
	M_\text{B}(U) = M(T) \, .
\eeq
The balance equation \eqref{energy_balance} can then be \textit{derived} from the mass-loss formula \eqref{mass_loss}.\footnote{Since Eq.~\eqref{energy_balance} is a functional equality, it can also be written as $\ud E / \ud t = - \mathcal{F}(t)$, where $t$ is the harmonic-coordinate time in the near-zone of the source.} Conversely, if the PN energy balance is assumed to be valid, then the identification \eqref{Mb_E} must hold. The Carter-Penrose diagram depicted in Fig.~\ref{fig:cons_diss} illustrates the previous discussion.

\begin{figure}[t!]
	\includegraphics[scale=0.3]{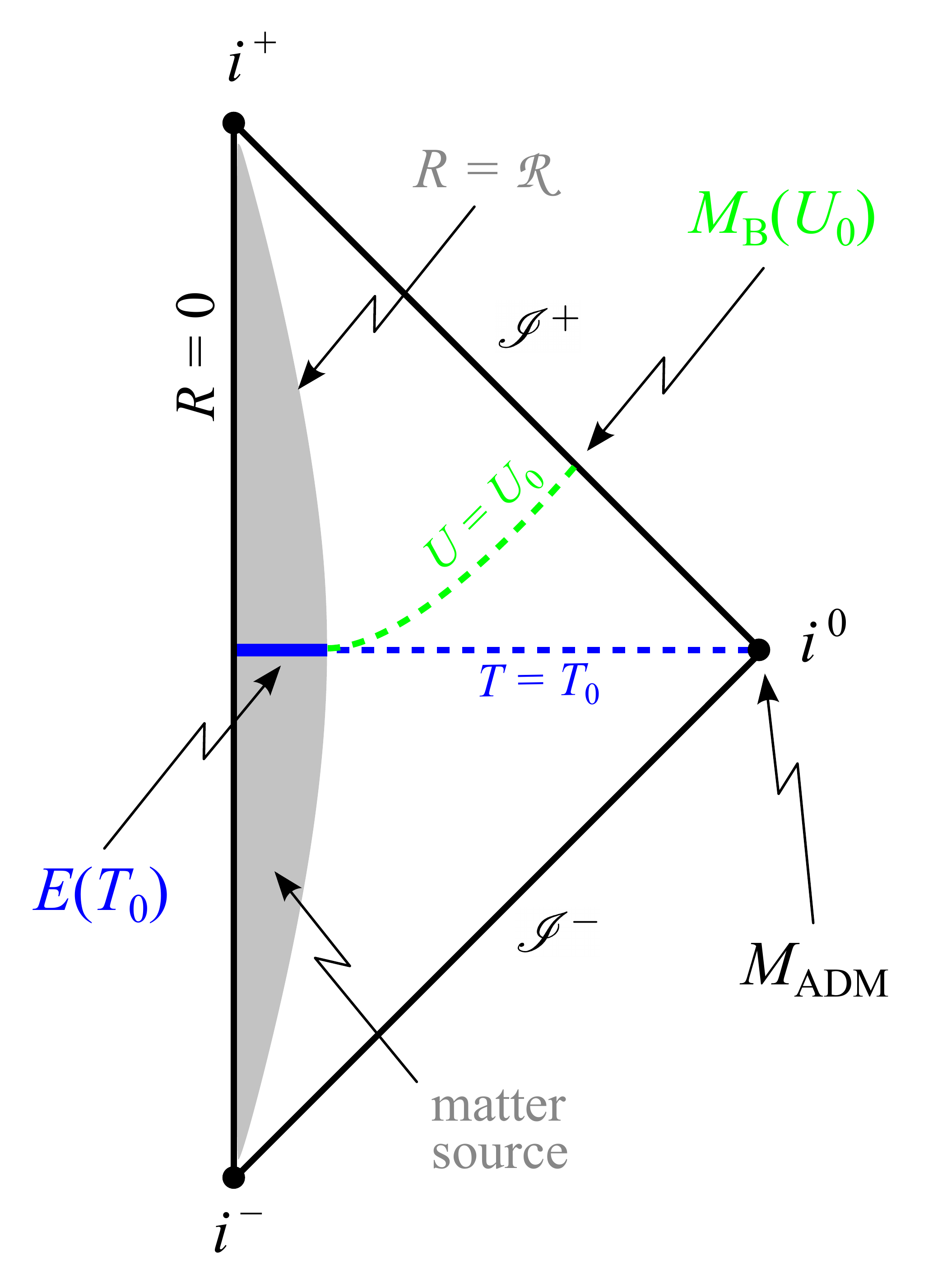}
	\caption{\footnotesize A non-radiative PN spacetime containing an ever-lasting gravitationally bound isolated matter source of constant size $\mathcal{R}$. The ADM mass $M_\text{ADM}$ coincides with the binding energy $M(T)=m+E(T)$, which is itself equal to the Bondi mass $M_\text{B}(U)$, at all times.}
	\label{fig:cons}
\end{figure}

While in the exact theory it is not possible to unambiguously split the conservative and dissipative parts of the orbital dynamics (\textit{e.g.} in the binary black hole spacetimes simulated in numerical relativity), this can be done in approximation methods such as PN expansions and black hole perturbation theory. For example, in post-Newtonian theory, one can discard the purely dissipative 2.5PN and 3.5PN radiation-reaction terms, and consider only the conservative dynamics at the Newtonian, 1PN, 2PN and 3PN orders. Since there is a one-to-one correspondance between the near-zone radiation-reaction force affecting the dynamics of the source and the fluxes of energy, linear momentum, and angular momentum computed from the wave-zone radiation field (see \textit{e.g.} Refs.~\cite{IyWi.93,IyWi.95}), this amounts to considering \textit{non-radiative} spacetimes. From Eq.~\eqref{Madm_Mb} --- in which we set $\mathcal{F} = 0$ --- and the identification \eqref{Mb_E}, we find that for such PN spacetimes, all the notions of mass that we have considered so far coincide, namely
\beq\label{Madm_Mb_E}
	M_\text{ADM} = M_\text{B} = M = \text{const}\,.
\eeq
See Fig.~\ref{fig:cons} for an illustration. We shall also assume that a similar result holds for the angular momenta, \textit{i.e.} that the angular momentum $J^i$ computed in PN theory from the near-zone metric coincides with the total angular momentum of the system, defined at spatial infinity.

In Sec.~\ref{sec:PNfirstlaw}, the first law of binary point-particle mechanics, as expressed by Eq.~\eqref{first_law} and its first integral \eqref{first_integral}, was derived directly from PN calculations, under the assumptions of helical symmetry and asymptotic flatness. When considering only the conservative part of the dynamics of a point-particle binary system on a circular orbit, both conditions are fulfilled by the post-Newtonian metric. For such orbits, the PN results in Sec.~\ref{sec:PNfirstlaw} establish algebraically that, for $M=m+E$, $J$ and $z_{1,2}$ given there, the first law is indeed satisfied up to the PN order involved.  We shall now show that this first law actually holds more generally so that, for the system being considered, it can be expected to be satisfied at all (conservative) PN orders.

\section{The first law of binary black hole mechanics}\label{sec:first_law}

Friedman, Ury{\=u} and Shibata~\cite{Fr.al.02} considered a one-parameter family of solutions of the Einstein field equations, describing an arbitrary number of black holes with a generic distribution of perfect fluid matter sources having compact support.\footnote{See Ref.~\cite{Ur.al.10} for the recent generalization to the magnetohydrodynamical case.} A perfect fluid is entirely characterized by its four-velocity $u^\alpha$, normalized to $g_{\alpha \beta} u^\alpha u^\beta = -1$, its energy density $\varepsilon(\rho,s)$, and pressure $P(\rho,s)$, both functions of the entropy per unit baryon mass $s$ and the conserved proper mass density $\rho$, such that $\nabla_\alpha (\rho u^\alpha) = 0$.

Each member of the family of geometries $\{ g_{\alpha\beta}(\lambda), u^\alpha(\lambda), \rho(\lambda), s(\lambda) \}$ is assumed to have a \textit{globally} defined Killing vector field $K^\alpha$. The Noether current associated with $K^\alpha$ assigns to each spacetime a conserved charge $Q$. The main result established in Ref.~\cite{Fr.al.02} relates the Eulerian variation $\delta Q \equiv \ud Q / \ud \lambda \vert_{\lambda=0}$ of the conserved charge $Q$ to the Eulerian variations $\delta A_n$ of the horizon surface areas $A_n$ of the black holes, as well as to the Lagrangian variations $\Delta (\ud M_\text{b})$, $\Delta (\ud S)$ and $\Delta (\ud C_\alpha)$ of the fluid's baryonic mass, entropy, and vorticity. This generalized first law explicitly reads
\beq\label{deltaQ}
	\delta Q = \int_\Sigma \left[ \bar{\mu} \, \Delta (\ud M_\text{b}) + \bar{T} \, \Delta (\ud S) + v^\alpha \Delta (\ud C_\alpha) \right] + \sum_n \frac{\kappa_n}{8\pi} \, \delta A_n \, ,
\eeq
where $\kappa_n$ are the uniform surface gravities of the black holes, $\bar{T} = T / u^t$ is the redshifted temperature, and $\bar{\mu} = (h - T s) / u^t$ the redshifted chemical potential (or specific Gibbs free energy), with $h = (\varepsilon + P) / \rho $ the specific enthalpy of the fluid. The spacelike velocity field $v^\alpha$ measures the peculiar velocities of the fluid elements with respect to the Killing vector field $K^\alpha$, whose integral curves define some preferred worldlines; it is defined by $u^\alpha = u^t (K^\alpha + v^\alpha)$, with $v^\alpha \nabla_\alpha t = 0$. The matter fields to be varied are given by
\begin{subequations}
	\begin{align}
		\ud M_\text{b} &\equiv \rho u^\alpha \ud \Sigma_\alpha \, , \label{dM} \\
		\ud S &\equiv s \, \ud M_\text{b} \, , \\
		\ud C_\alpha &\equiv h u_\alpha \ud M_\text{b} \, ,
	\end{align}
\end{subequations}
with $\ud \Sigma_\alpha = - n_\alpha \sqrt{\gamma} \, \ud^3 x$ the volume element on the spacelike ($t = \text{const}$) hypersurface $\Sigma$, covered by Cartesian coordinates $\{x^i\}$, and over which the integral in Eq.~\eqref{deltaQ} is performed; $n^\alpha$ is the future-pointing unit timelike vector normal to $\Sigma$, and $\gamma = \text{det}(\gamma_{ij})$ the determinant of the induced metric $\gamma_{ij}$ on $\Sigma$.

\subsection{The first law for point-particle binaries on circular orbits}
\label{subsec:proof}

We are interested in applying the general result \eqref{deltaQ} to the particular case of a compact binary system, whose components move on exactly circular orbits. For such spacetimes, the geometry is invariant along the direction of the helical Killing vector $K^\alpha = t^\alpha + \Omega \, \phi^\alpha$, where $\Omega$ is a constant, which is identified with the angular frequency $\ud \varphi / \ud t$ of the circular motion; the vectors $t^\alpha \equiv {(\partial_t)}^\alpha$ and $\phi^\alpha \equiv {(\partial_\phi)}^\alpha$ are part of the natural basis $(\partial_t,\partial_r,\partial_\theta,\partial_\phi)$, with $\{r,\theta,\phi\}$ the spherical coordinate system associated with the Cartesian coordinates $\{x^i\}$ in the usual way, \textit{i.e.} $x^1 = r \sin{\theta} \cos{\phi}$, $x^2 = r \sin{\theta} \sin{\phi}$ and $x^3 = r \cos{\theta}$.

For asymptotically flat spacetimes, the basis vectors $t^\alpha$ and $\phi^\alpha$ are both asymptotic Killing vectors, reflecting the invariance of Minkowski spacetime under time translations and spatial rotations. Then, the variation $\delta Q$ of the conserved charge $Q$ associated with the helical Killing vector $K^\alpha$ is given by~\cite{Fr.al.02}
\beq\label{deltaQbis}
	\delta Q = \delta M - \Omega \, \delta J \, ,
\eeq
where $M$ is the ADM mass, and $J$ the norm of the total angular momentum $J_i$ of the system, defined as a surface integral at spatial infinity, in terms of the three rotational Killing vectors of the flat metric $\eta_{\alpha\beta}$ (\textit{cf.} Eq.~(109) of Ref.~\cite{Fr.al.02}). The variational equation \eqref{deltaQbis} holds in the center-of-mass frame, in which the ADM three-momentum $P_i$ vanishes.

As discussed in the Introduction, we shall model the compact objects (namely non spinning black holes and/or neutron stars) as two point particles with ``Schwarzschild'' masses $m_1$ and $m_2$. We therefore consider the simple case of a perfect fluid with vanishing pressure $P$, temperature $T$, specific entropy $s$, and peculiar velocity $v^\alpha$. The redshifted chemical potential then simply reads $\bar{\mu} = 1 / u^t$. Combining Eq.~\eqref{deltaQbis} with \eqref{deltaQ}, in which we set $\kappa_n \, \delta A_n = 0$, the generalized first law reduces to
\beq\label{deltaM_OmegaJ}
	\delta M - \Omega \, \delta J = \int_\Sigma z \, \Delta (\ud M_\text{b}) \, ,
\eeq
where $z \equiv 1 / u^t$ is the redshift factor. Now, in order to evaluate the Lagrangian variation $\Delta (\ud M_\text{b})$ of the baryonic mass element \eqref{dM}, we use the explicit expression of the proper mass density $\rho$ for two point masses, namely
\beq
	\rho(\bm{x},t) = \frac{1}{\sqrt{-g}} \, \sum_{A=1}^2 m_A \, z_A \, \delta[\bm{x}-\bm{y}_A(t)] \, ,
\eeq
where $\bm{y}_A(t)$ are the coordinate trajectories of the particles ($A=1,2$), $\delta$ is the usual three-dimensional Dirac distribution, such that $\int \ud^3 x \, \delta(\bm{x}) = 1 $, and $g = \text{det}(g_{\alpha \beta})$ is the determinant of the covariant metric $g_{\alpha \beta}$. In a $3+1$ decomposition, we have $\sqrt{-g} = N \sqrt{\gamma}$, where $N$ is the lapse function. Since the four-velocities $u_A^\alpha$ of the particles are tangent to the HKV evaluated at their coordinate locations, \textit{i.e.} $u_A^\alpha = u_A^t \, K^\alpha(\bm{y}_A,t)$, we find that \eqref{dM} explicitly reads
\beq\label{dMbis}
	\ud M_\text{b} = - \ud^3 x \, \sum_{A=1}^2 m_A \, \frac{{(Kn)}_A}{N_A} \, \delta(\bm{x}-\bm{y}_A) \, .
\eeq
We have introduced the shorthands $N_A \equiv N(\bm{y}_A,t)$ and $(Kn) \equiv K_\alpha n^\alpha$. From the 3+1 decomposition of the time evolution vector $t^\alpha$ in terms of the lapse $N$ and shift $N^\alpha$, namely $t^\alpha = N n^\alpha + N^\alpha$, the HKV reads $K^\alpha = N n^\alpha + (N^\alpha + \Omega \phi^\alpha)$. Since the shift vector $N^\alpha$ and the basis vector $\phi^\alpha$ are both tangent to $\Sigma$, we get $K_\alpha n^\alpha = - N$, which in turn implies ${(Kn)}_A = - N_A$ at the location of each particle. This simplifies the expression of the baryonic mass element \eqref{dMbis}\footnote{Namely, $\ud M_\text{b} = \rho_* \,\ud^3 x$ where $\rho_* = \sum_{A} m_A \, \delta(\bm{x}-\bm{y}_A)$ denotes the baryonic coordinate density.} which, once substituted into \eqref{deltaM_OmegaJ}, yields our final result in the form\footnote{For clarity and uniformity in notation, we have made the substitution $\Delta m_A \rightarrow \delta m_A$ for the Lagrangian variations of the individual masses $m_1$ and $m_2$.}
\beq\label{Smarr_2pp_delta}
	\delta M - \Omega \, \delta J = z_1 \, \delta m_1 + z_2 \, \delta m_2 \, ,
\eeq
which we recognize as our first law of binary point-particle mechanics \eqref{first_law} derived using post-Newtonian theory. This first law compares two neighbouring helically symmetric, asymptotically flat solutions of the field equations, and tells how the changes in the ``baryonic'' masses of the bodies will affect the ADM mass and angular momentum of the binary system. Alternatively, the first law of binary point-particle mechanics could probably be derived in the Hamiltonian approach by varying the individual degrees of freedom of the point masses, namely their positions, momenta and masses. Some consequences of Eq.~\eqref{Smarr_2pp_delta} have already been explored, notably the existence of the first integral given by
\beq\label{Smarr_2pp}
	M - 2 \Omega J = m_1 z_1 + m_2 z_2 \, .
\eeq
In Sec.~\ref{sec:PNfirstlaw} we derived the first law \eqref{Smarr_2pp_delta} directly from post-Newtonian calculations. Here we recovered this important relation based on first principles in general relativity; in particular our derivation does not rely on a post-Newtonian expansion. Hence we expect the first law \eqref{Smarr_2pp_delta}, as well as all of its consequences, including Eq.~\eqref{Smarr_2pp}, to hold at \textit{all} conservative orders in a PN expansion.

\subsection{Analogies with single and binary black hole mechanics}
\label{subsec:analogies}

In this Section we point out some interesting analogies between the result \eqref{Smarr_2pp_delta} and its consequence \eqref{Smarr_2pp} for point-particle binaries on one hand, and some well-known relations regarding the mechanics (or thermodynamics) of single or binary black holes on the other hand. In the particular case of an asymptotically flat vacuum spacetime with two black holes on quasi-circular orbits, the general result \eqref{deltaQ} reduces to\footnote{See Sec.~\ref{sec:introduction} for a discussion of helical symmetry and asymptotic flatness in binary black hole spacetimes.}
\beq\label{Smarr_2BH_delta}
	\delta M - \Omega \, \delta J = \kappa_1 \, \frac{\delta A_1}{8\pi} + \kappa_2 \, \frac{\delta A_2}{8\pi} \, .
\eeq
This variational relation can be viewed as a generalization to the binary black hole case of the celebrated first law of black hole mechanics $\delta M - \Omega_\text{H} \, \delta J = \kappa \, \delta A / (8\pi)$, which holds for \textit{any} non-singular, asymptotically flat perturbation of a stationary and axisymmetric black hole of mass $M$, intrinsic angular momentum (or spin) $J \equiv M a$, surface area $A$, uniform surface gravity $\kappa$, and angular frequency $\Omega_\text{H}$ on the horizon~\cite{Ba.al.73,Wa.93}. In the binary black hole case, the horizon angular velocity $\Omega_\text{H}$ of a single rotating black hole is replaced by the orbital frequency $\Omega$ of the binary.

The surface area $A$ of a black hole is related to its irreducible (or Christodoulou) mass $m_\text{irr}$ through $m^2_\text{irr} = A / (16\pi)$~\cite{Ch.70,ChRu.71}. We may thus substitute $\kappa \, \delta A / (8\pi) \longrightarrow (4 m_\text{irr} \,\kappa) \, \delta m_\text{irr}$ in both terms in the RHS of Eq.~\eqref{Smarr_2BH_delta}. Then, comparing the first law of binary black hole mechanics \eqref{Smarr_2BH_delta} to Eq.~\eqref{Smarr_2pp_delta} for the analogous binary point-particle case, we notice the formal analogies
\beq\label{analogies}
	m \longleftrightarrow m_\text{irr} \, , \quad z \longleftrightarrow 4 m_\text{irr} \kappa \, .
\eeq
Both analogies are rather intuitive and physically appealing. Indeed, one might expect the irreducible mass $m_\text{irr}$ of a non-rotating black hole to be analogous to the ``post-Newtonian'' mass $m$ of a point particle.\footnote{Notice however that the assumption of helical symmetry, which is necessary to derive the first law \eqref{Smarr_2BH_delta}, requires that the two black holes are in co-rotation, and must therefore have non-zero spins~\cite{Go.al.02,Fr.al.02}. It would be interesting to generalize the binary point-particle first law \eqref{Smarr_2pp_delta} by including spin effects, \textit{e.g.} using a pole-dipole model for the spinning point masses. For rotating black holes, the post-Newtonian mass $m$ should be identified with the total mass of the black hole, including the effect of the spin $S$; hence $m^2 = m_\text{irr}^2+S^2/(4m_\text{irr}^2)$. See \textit{e.g.} Ref.~\cite{Bl.02}.} Furthermore, the surface gravity $\kappa$ of a black hole may naturally be related to the redshift $z$ of the ``associated'' point mass, \textit{via} the gravitational redshift, or Einstein effect (\textit{cf.} Detweiler's Gedanken experiment in Sec.~II~C of~\cite{De.08}). In particular, for two Schwarzschild black holes on quasi-circular orbits, but far enough apart so that they can be viewed --- in first approximation --- as isolated, the surface gravity $\kappa \longrightarrow (4m_\text{irr})^{-1}$, while for the point particles we have $z \longrightarrow 1$. Therefore, in that limiting case the analogies \eqref{analogies} seem perfectly sound.

Then, following exactly the same steps as in Sec.~\ref{sec:PNfirstlaw}, but with the irreducible masses of the black holes playing the role of the post-Newtonian masses of the point particles, it can easily be established that
\beq\label{Smarr_2BH}
	M - 2 \Omega J = \kappa_1 \, \frac{A_1}{4\pi} + \kappa_2 \, \frac{A_2}{4\pi} \, .
\eeq
This last relation can be viewed as a generalization to the binary black hole case of Smarr's formula $M - 2 \Omega_\text{H} J = \kappa A / (4 \pi)$ for a single Kerr black hole~\cite{Sm.73}. The RHS side of Eq.~\eqref{Smarr_2BH} is the sum of two terms of the form $\kappa A / (4 \pi) = 4 m_\text{irr}^2 \kappa$; one for each black hole. This binary black hole expression should be compared to the RHS of Eq.~\eqref{Smarr_2pp} for the point-particle case. We then find the analogy
\beq
	m z \longleftrightarrow 4 m_\text{irr}^2 \kappa \, ,
\eeq
\textit{i.e.} precisely what is expected from the analogies \eqref{analogies}, which were drawn from the differential relations \eqref{Smarr_2pp_delta} and \eqref{Smarr_2BH_delta}.

It would be interesting to investigate how far the formal analogies \eqref{analogies} can be pushed, especially in the relevant regime where the two black holes or point masses are in co-rotation and interact strongly with each other, namely when $z \neq 1$. One might for example try to recover the point-particle results \eqref{Smarr_2pp_delta} and \eqref{Smarr_2pp} starting from the black hole results \eqref{Smarr_2BH_delta} and \eqref{Smarr_2BH}, by taking some suitable point-particle limit. In that respect, one might use the method proposed in Ref.~\cite{GrWa.08} to derive the gravitational self-force~\cite{Mi.al.97,QuWa.97} for the motion of a point particle in some background curved spacetime, by scaling down the mass $M$ and size $R$ of an extended compact object, while helding its compactness $M/R$ fixed. Such an analysis might improve our understanding of the validity of modelling extended objects such as black holes by point masses in general relativity. This would be valuable because both post-Newtonian methods and gravitational self-force calculations rely heavily on such idealizations~\cite{Bl.06,Po.al.11}. These questions should be addressed in future work.

Finally, one can also consider the mixed case of a circular-orbit compact binary system composed of a black hole and a point particle. This corresponds to the model usually adopted in black hole perturbation theory and the gravitational self-force. Calculations similar to those detailed above naturally yield the results
\begin{subequations}
	\begin{align}
		\delta M - \Omega \, \delta J &= \kappa \, \frac{\delta A}{8\pi} + z \, \delta m \, , \\
		M - 2 \Omega J &= \kappa \, \frac{A}{4\pi} + m z \, .
	\end{align}
\end{subequations}

\subsection{Alternative derivation of the first integral relations}
\label{sec:Komar}

In this Section we show that the algebraic first integral relations \eqref{Smarr_2pp} and \eqref{Smarr_2BH} can also be obtained by standard techniques derived from the definition of the Komar mass and angular momentum (see \textit{e.g.}~\cite{Wal}). We introduce a Cartesian-type coordinate system $\{t,x^i\}$, as well as the associated spherical coordinate system $\{t,r,\theta,\phi \}$. These coordinates are chosen such that the metric $g_{\alpha\beta}$ is explicitly asymptotically Minkowskian at spatial infinity, \textit{i.e.} $g_{\alpha\beta} \to \eta_{\alpha\beta}$ when $r \to +\infty$. Performing a 3+1 decomposition of the four-dimensional spacetime, the metric reads
\beq\label{3+1}
	\ud s^2 = - N^2 \ud t^2 + \gamma_{ij} \left( \ud x^i + N^i \ud t \right) \! \left( \ud x^j + N^j \ud t \right) ,
\eeq
where $N$ is the lapse function, $N^i$ the shift vector, and $\gamma_{ij}$ the three-metric of $t = \text{const}$ hypersurfaces. Let $\Sigma$ be one such spacelike hypersurface, and $S = \partial \Sigma$ be the two-sphere at spatial infinity representing its boundary.

\begin{figure}[t!]
	\includegraphics{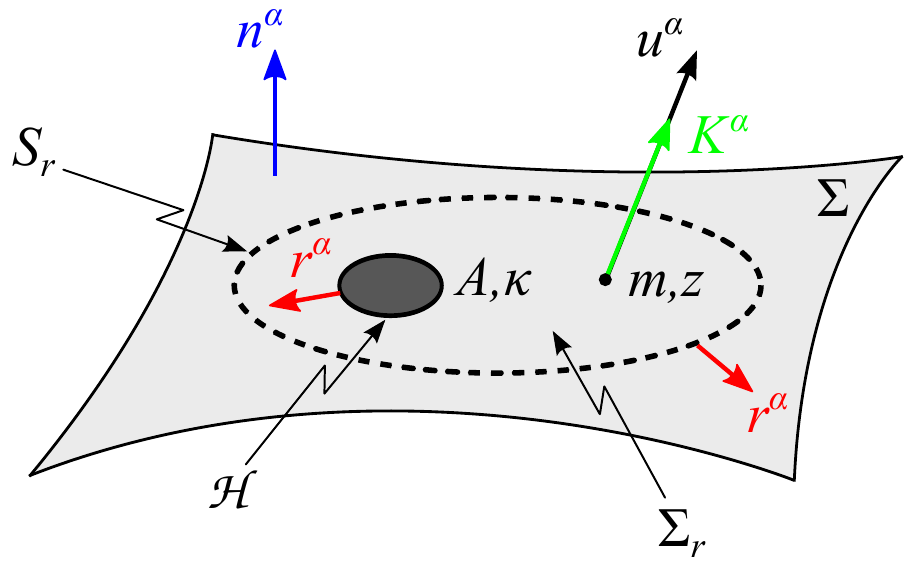}
	\caption{\footnotesize A spatial slice $\Sigma$ of a circular-orbit compact binary spacetime. The black hole is characterized by its horizon area $A$ and uniform surface gravity $\kappa$, while the point particle has a mass $m$ and redshift factor $z = 1 / u^t$, and is such that its four-velocity reads $u^\alpha = u^t K^\alpha$.}
	\label{fig:3+1}
\end{figure}

The notions of  Komar mass $M_\text{K}$ and Komar angular momentum $J_\text{K}$ are commonly introduced for stationary and axisymmetric spacetimes~\cite{Ko.59}. Conserved currents can be built from the Killing vectors associated with these symmetries, leading to some well-defined notions of mass and angular momentum defined as surface integrals over these currents. However, the Komar quantities can also be defined for more generic spacetimes, which are neither stationary, nor axisymmetric, as long as they are asymptotically flat. For such geometries, the Komar mass and angular momentum are defined as 
\begin{subequations}\label{komar}
	\begin{align}
		M_\text{K} &\equiv - \frac{1}{8\pi} \lim_{r \rightarrow \infty} \oint_{S_r} \! \nabla^\alpha t^\beta \, \ud S_{\alpha \beta} \, , \label{Mkomar} \\
		J_\text{K} &\equiv \frac{1}{16\pi} \lim_{r \rightarrow \infty} \oint_{S_r} \! \nabla^\alpha \phi^\beta \, \ud S_{\alpha \beta} \, , \label{Jkomar}
	\end{align}
\end{subequations}
where the basis vectors $t^\alpha \equiv {(\partial_t)}^\alpha$ and $\phi^\alpha \equiv {(\partial_\phi)}^\alpha$ are \textit{asymptotic} Killing vectors, reflecting the invariance of Minkowski spacetime under time translations and spatial rotations. The integrals in Eqs.~\eqref{komar} are performed over a two-sphere $S_r$ of coordinate radius $r$, which goes to the two-sphere at spatial infinity $S$ as the limit $r \to +\infty$ is taken at the end of the calculation.\footnote{For stationary and/or axisymmetric spacetimes, $t^\alpha$ and/or $\phi^\alpha$ are Killing vectors on the entire spacetime manifold, and the integrals may as well be performed over any spacelike topological two-sphere enclosing the matter source (where the matter stress-energy tensor $T^{\alpha\beta} \neq 0$).} The two-sphere $S_r$ is the boundary of an hypersurface $\Sigma_r$, which is itself part of the hypersurface $\Sigma$ bounded by $S$. The surface element on $S_r$ reads $\ud S_{\alpha \beta} = 2 r_{[\alpha} n_{\beta]} \sqrt{\sigma} \, \ud^2 y$, where $n^\alpha$ is the future-pointing unit timelike vector normal to $\Sigma$, and $r^\alpha$ the outward-pointing unit spacelike vector normal to $S_r$, and tangent to $\Sigma$. The metric induced on $\Sigma$ is $\gamma_{\alpha\beta}=g_{\alpha\beta}+n_\alpha n_\beta$, and the metric induced on $S_r$ is $\sigma_{\alpha\beta}=\gamma_{\alpha\beta}-r_\alpha r_\beta$, with determinant $\sigma = \text{det}(\sigma_{ab})$ in coordinates $\{y^a\}$ covering $S_r$ (see Fig.~\ref{fig:3+1} for notations). Note that the integral \eqref{Jkomar} for the angular momentum $J_\text{K}$ exhibits the famous ``Komar anomalous factor'' $-1/2$ with respect to the integral \eqref{Mkomar} for the mass $M_\text{K}$~\cite{Ka.85}.

Although $t^\alpha$ and $\phi^\alpha$ are only asymptotic Killing vectors, our assumption of helical symmetry implies that $K^\alpha = t^\alpha + \Omega \, \phi^\alpha$ is a Killing vector field over the entire spacetime. The interesting combination of $M_\text{K}$ and $J_\text{K}$ is therefore given by
\beq\label{surf_int}
	M_\text{K} - 2 \Omega J_\text{K} = - \frac{1}{8\pi} \oint_S \nabla^\alpha K^\beta \, \ud S_{\alpha \beta} \, .
\eeq
Because of Killing's equation $\nabla_{(\alpha} K_{\beta)} = 0$, the tensor $\nabla^\alpha K^\beta$ is antisymmetric. Using a version of the Stokes theorem for rank $2$ antisymmetric tensor fields, the surface integral in Eq.~\eqref{surf_int} can be converted into a volume integral over the hypersurface $\Sigma$ bounded by $S$ (and the multiple event horizons $\mathcal{H}_n$ for spacetimes with black holes):
\beq
	\oint_{S} \nabla^\alpha K^\beta \, \ud S_{\alpha \beta} = \sum_n \oint_{\mathcal{H}_n} \!\! \nabla^\alpha K^\beta \, \ud S_{\alpha \beta} + 2 \int_\Sigma \nabla_\beta (\nabla^\alpha K^\beta) \, \ud \Sigma_\alpha \, .
\eeq
The volume element on $\Sigma$ reads $\ud \Sigma_\alpha = - n_\alpha \sqrt{\gamma} \, \ud^3 x$, with $\gamma = \text{det}(\gamma_{ij})$ the determinant of the spatial metric $\gamma_{ij}$. Then, Killing's equation together with the non-commutation of covariant derivatives yields the well-known formula $\nabla_\beta \nabla^\alpha K^\beta = R^\alpha_{\phantom{\alpha}\beta} \, K^\beta$. Using the Einstein equation to replace the Ricci tensor $R_{\alpha\beta}$ in favor of the stress-energy tensor $T_{\alpha\beta}$, we obtain
\beq\label{int}
	M_\text{K} - 2 \Omega J_\text{K} = - \frac{1}{8\pi} \sum_n \oint_{\mathcal{H}_n} \!\! \nabla^\alpha K^\beta \, \ud S_{\alpha \beta} + 2 \int_\Sigma \left( T_{\alpha\beta} - \frac{1}{2} T g_{\alpha\beta} \right) n^\alpha K^\beta \sqrt{\gamma} \, \ud^3 x \, ,
\eeq
where $T \equiv g_{\alpha\beta} T^{\alpha\beta}$. Similar expressions for $M_\text{K}$ and $J_\text{K}$ (separately) can be found in textbooks, for stationary and/or axisymmetric spacetimes; see \textit{e.g.} Refs.~\cite{Wal,Poi,Go.07}. We shall now successively apply the general result \eqref{int} to the particular cases of binary point particle and binary black hole spacetimes.

\subsubsection{Binary point-particle case}

We consider first a spacetime with two point masses modelling two compact objects on a circular orbit; hence we discard the first term in the RHS of Eq.~\eqref{int}. The stress-energy tensor is generated by two massive worldlines, and reads

\beq\label{Tmunu}
	T^{\alpha\beta}(\bm{x},t) = \frac{1}{\sqrt{-g}} \, \sum_{A=1}^2 m_A \, z_A \, u_A^\alpha u_A^\beta \, \delta[\bm{x}-\bm{y}_A(t)] \, ,
\eeq
where $\bm{y}_A(t)$ are the coordinate trajectories of the particles ($A=1,2$), with normalized four-velocities $u_A^\alpha$, and where $z_A = 1/u_A^t$ are the redshift variables. The determinants $g = \text{det}(g_{\alpha\beta})$ and $\gamma = \text{det}(\gamma_{ij})$ are related by $\sqrt{-g} = N \sqrt{\gamma}$. Now, remember that the four-velocities $u_A^\alpha$ of the particles are tangent to the helical Killing vector $K^\alpha$ evaluated at their coordinate locations; hence
\beq\label{uk}
	u_A^\alpha = u_A^t \, K_A^\alpha \, ,
\eeq
where we introduced the shorthand $K_A^\alpha \equiv K^\alpha(\bm{y}_A,t)$ (\textit{cf.} Fig.~\ref{fig:3+1}). Inserting \eqref{Tmunu} into \eqref{int}, contracting the tensors, and making use of Eq.~\eqref{uk}, we obtain
\beq\label{combiKomar}
	M_\text{K} - 2 \Omega J_\text{K} = - \sum_{A=1}^2 m_A z_A \, \frac{{(Kn)}_A}{N_A} \, ,
\eeq
where we recall that $(Kn) \equiv K_\alpha n^\alpha$. As we have already proved below Eq.~\eqref{dMbis} in Sec.~\ref{subsec:proof} above, we have ${(Kn)}_A = - N_A$ at the coordinate locations of the particles. We thus recover an expression which is formally identical to our relation \eqref{Smarr_2pp}, with the Komar quantities playing the role of the ADM mass and angular momentum, namely
\beq
M_\text{K} - 2 \Omega J_\text{K} = m_1 z_1 + m_2 z_2 \, .
\eeq

Finally, it can be shown that if the foliation ${(\Sigma_t)}_{t \in \mathbb{R}}$ is adapted to axisymmetry at spatial infinity (namely $n^\alpha \phi_\alpha \to 0$ when $r \to +\infty$), then the Komar angular momentum \eqref{Jkomar} coincides with the total angular momentum~\cite{Go.07}: $J_\text{K} = J$. On the other hand, for stationary, asymptotically flat spacetimes, if the foliation is such that the unit normal $n^\alpha$ coincides with the timelike Killing vector $t^\alpha$ at spatial infinity (\textit{i.e.} $N \to 1$ and $N^i \to 0$ when $r \to +\infty$), then the Komar mass \eqref{Mkomar} coincides with the ADM mass~\cite{Be.78,AsMa2.79}. In Ref.~\cite{GoBo.94}, the equality $M_\text{K} = M$ allowed the definition of a relativistic version of the classical virial theorem for stationary, asymptotically flat spacetimes. More recently, Shibata \textit{et al.}~\cite{Sh.al.04} have shown that the equality $M_\text{K} = M$ holds for a much larger class of spacetimes (in particular they could relax the hypothesis of stationarity); see Eqs.~(3.7)--(3.9) in Ref.~\cite{Sh.al.04} for the required asymptotic conditions on the lapse $N$, shift $N^i$, three-metric $\gamma_{ij}$, extrinsic curvature $K_{ij}$, and their spatial derivatives. Assuming that our non-radiative, helically symmetric, post-Newtonian spacetimes would satisfy these falloff conditions, we then have
\beq\label{Komar_ADM}
	M_\text{K} - 2 \Omega J_\text{K} = M - 2 \Omega J \, ,
\eeq
and we thus recover the algebraic first integral relation \eqref{Smarr_2pp}. From this one sees that the curious factor of 2 in Eq.~\eqref{Smarr_2pp} is actually related to Komar's anomalous factor \cite{Ka.85}.

\subsubsection{Binary black hole case}

We turn next to the case of a vacuum spacetime with two co-rotating black holes on quasi-circular orbits. From the general result \eqref{int}, in which we set $T_{\alpha\beta} = 0$, we find
\beq\label{bbh_case}
	M_\text{K} - 2 \Omega J_\text{K} = - \frac{1}{8\pi} \oint_{\mathcal{H}_1} \!\! \nabla_\alpha K_\beta \, \ud S^{\alpha \beta} - \frac{1}{8\pi} \oint_{\mathcal{H}_2} \!\! \nabla_\alpha K_\beta \, \ud S^{\alpha \beta} \, .
\eeq
Following Friedman \textit{et al.}~\cite{Fr.al.02}, we write the surface two-form as $\ud S^{\alpha\beta} = 2 K^{[\alpha} \ell^{\beta]} \sqrt{\sigma} \, \ud^2 y$, where $\ell^\alpha$ is the unique null vector orthogonal to $\mathcal{H}_n$ and such that $K^\alpha \ell_\alpha = -1$. For two black holes in co-rotation, the helical Killing vector $K^\alpha$ is tangent to the null generators of the horizons; hence the surface gravity $\kappa$ is defined by the usual relation $K^\beta \nabla_\beta K^\alpha = \kappa \, K^\alpha$. The integrand in Eq.~\eqref{bbh_case} thus reduces to $\nabla_\alpha K_\beta \, \ud S^{\alpha \beta} = - 2 \kappa \sqrt{\sigma} \, \ud^2 y$, yielding
\beq
	M_\text{K} - 2 \Omega J_\text{K} = \frac{1}{4\pi} \oint_{\mathcal{H}_1} \!\! \kappa \, \ud A + \frac{1}{4\pi} \oint_{\mathcal{H}_2} \!\! \kappa \, \ud A \, ,
\eeq 
with $\ud A = \sqrt{\sigma} \, \ud^2 y$ the surface element. For a single black hole, the zeroth law of black hole mechanics states that $\kappa$ is uniform over the event horizon \cite{Ba.al.73}. This result was generalized to the binary black hole case in Ref.~\cite{Fr.al.02}; hence $\kappa$ can be pulled out of the integrals. Combining this with the argument \eqref{Komar_ADM}, we recover the result \eqref{Smarr_2BH}.

On the other hand, several authors have previously established a relation similar to \eqref{Smarr_2BH}, in the case of two co-rotating black holes on quasi-circular orbits, namely~\cite{De.89,Go.al.02}
\beq\label{Smarr_2BH_bis}
	M - 2 \Omega J = \frac{1}{4\pi} \oint_{\mathcal{H}_1} \!\! \nabla_\alpha N \, \ud S^\alpha + \frac{1}{4\pi} \oint_{\mathcal{H}_2} \!\! \nabla_\alpha N \, \ud S^\alpha \, ,
\eeq
where $\ud S^\alpha = r^\alpha \, \ud A$. We shall now derive this result from Eqs.~\eqref{Komar_ADM}--\eqref{bbh_case}, thus establishing that for such binary black hole spacetimes, the radial projection of the gradient of the lapse, $r^\alpha \nabla_\alpha N$, essentially coincides with the surface gravity $\kappa$ on each horizon, in agreement with physical intuition. We assume that the conditions for which \eqref{Komar_ADM} holds are met. Making use of the antisymmetry of the tensor $\nabla_\alpha K_\beta$, we have
\beq\label{extr_curv}
	\nabla_\alpha K_\beta \, \ud S^{\alpha \beta} = 2 n^\beta \nabla_\alpha K_\beta \, \ud S^\alpha = - 2 \nabla_\alpha N \, \ud S^\alpha - 2  K^\beta \nabla_\alpha n_\beta \, \ud S^\alpha \, ,
\eeq
where we performed an integration by parts and used the $3+1$ decomposition $K^\alpha = N n^\alpha + m^\alpha$, with $m^\alpha = N^\alpha + \Omega \, \phi^\alpha$ the shift vector in a co-rotating frame. The second term in the RHS of Eq.~\eqref{extr_curv} can be expressed using the extrinsic curvature tensor $K_{\alpha\beta}$, \textit{via} the well-known relation $\nabla_\alpha n_\beta = - K_{\alpha \beta} - n_\alpha \nabla_\beta \ln{N}$. Next, we notice that $K_{\alpha\beta}$ is both spacelike and symmetric, and that $n_\alpha r^\alpha = 0$; hence Eq.~\eqref{bbh_case} becomes
\beq\label{int_alt}
	{M} - 2 \Omega {J} = \frac{1}{4\pi} \sum_{A=1}^2 \oint_{\mathcal{H}_A} \!\! \nabla_i N \, \ud S^i - \frac{1}{4\pi} \sum_{A=1}^2 \oint_{\mathcal{H}_A} \!\! m^i K_{ij}  \, \ud S^j \, .
\eeq
Finally, because the HKV field $K^\alpha = N n^\alpha + m^\alpha$ must become null on the event horizons of the black holes, we have $N^2 = \gamma_{ij} \, m^i m^j$ on $\mathcal{H}_1$ and $\mathcal{H}_2$. Following the authors of Refs.~\cite{De.89,Go.al.02}, we impose the usual boundary condition of vanishing lapse on the horizons, and thus find that $m^i = 0$ in the second term in the RHS of Eq.~\eqref{int_alt} ($\gamma_{ij}$ being positive-definite). We thus recover the relation \eqref{Smarr_2BH_bis} established in \cite{De.89,Go.al.02}, based on the 3+1 decomposition of the Einstein field equations, in the presence of a HKV.

\section{Applications of the first law}\label{sec:application}

In this Section we consider a number of practical applications of our first law for compact binaries. The first is an application when one mass is much smaller than the other, so that a perturbative analysis is applicable.  The second is an application in which numerical results from a perturbation analysis for $z_1$ can be combined with relationships like \eqref{j45}--\eqref{pq45} to determine previously unknown coefficients in the PN expressions of $E$ and $J$.

\subsection{New gauge-invariant energy and angular momentum}
\label{subsec:newgaugeinv}

\subsubsection{Implications of the first law for perturbation analysis}
\label{subsec:pert_inps}

Perturbation analysis is applicable in the ``two-body problem'' when the mass of one of the bodies is much smaller than the mass of the other. Typically for our purposes, the large mass body will be treated as a black hole of mass $m_2$, while the small mass body may be treated either as a point particle or as a small black hole of mass $m_1 \ll m_2$. In this context, we have the usual notions of perturbative energy per unit mass and angular momentum per unit mass of the particle $m_1$,
\begin{subequations}\label{EJinit}
	\begin{align}
		 {\cal E}_1 &\equiv - u_{1t}\, , \\
		 {\cal J}_1 &\equiv u_{1\varphi}\, ,
	\end{align}
\end{subequations}
where $u_{1t}$ and $u_{1\varphi}$ are covariant components of the particle's four-velocity $u_1^\alpha$. It was shown in Ref.~\cite{De.08} that $z_1=1/u_1^t={\cal E}_1-\Omega{\cal J}_1$. As Detweiler pointed out~\cite{De.08}, ${\cal E}_1$ and ${\cal J}_1$ are not separately gauge invariant, while the redshift observable $z_1$ is. In the notation of Ref.~\cite{De.08}, Eqs.~(22) and (23) given there are, to first order in the mass ratio $q \equiv m_1 / m_2$,
\begin{subequations}\label{det_22_23}
	\begin{align}
{\cal E}_1 &= \frac{r-2m_2}{\sqrt{r(r-3m_2)}}\bigg[1-\frac{\bar{u}^\alpha \bar{u}^\beta h_{\alpha\beta}}{2}-\frac{r}{4}\bar{u}_\alpha \bar{u}_\beta \frac{\partial h^{\alpha\beta}}{\partial r}\bigg] \, , \\
{\cal J}_1 &= \frac{r \sqrt{m_{2}}}{\sqrt{r-3m_2}}\bigg[1-\frac{\bar{u}^\alpha \bar{u}^\beta h_{\alpha\beta}}{2}-\frac{r(r-2m_2)}{4m_2}\bar{u}_\alpha \bar{u}_\beta \frac{\partial h^{\alpha\beta}}{\partial r}\bigg] \, ,
	\end{align}
\end{subequations}
in which use has been made of conditions \eqref{eqdot} and \eqref{eqddot} for quasi-circular orbits, and where $\bar{u}^\alpha$ is a convenience introduced by Detweiler ---  it does {\textit{not}} denote the four-velocity of the particle but is defined in \eqref{ubar} below; and $h_{\alpha\beta}$ is the regularized metric perturbation, which is a smooth vacuum solution in the neighborhood of the particle \cite{DeWh.03}.\footnote{In this Subsection, $h_{\alpha\beta}$ is indeed a regular metric perturbation in the Schwarzschild geometry, and not the PN metric perturbation from flat space used in Sec.~\ref{sec:PN}, as defined in the first footnote in Sec.~\ref{sec:log_contrib}.} Making use of Eq.~(28) in \cite{De.08} we obtain (see also Eq.~(2.7) from Paper I)
\beq
\frac{m_2}{r} = y + \frac{r(1-3y)}{6} \,\bar{u}^\alpha \bar{u}^\beta \frac{\partial h_{\alpha\beta}}{\partial r}\, ,
\label{m2overragain}
\eeq
where we have introduced the dimensionless invariant PN parameter $y \equiv (m_2 \Omega)^{2/3}$. It will also be convenient to define a coordinate invariant measure of the constant orbit separation, \textit{via} $r_\Omega \equiv m_2 / y$. Then, using \eqref{m2overragain}, Eqs.~\eqref{det_22_23} can be rewritten, to first order in $q$, as
\begin{subequations}\label{detEJ}
	\begin{align}
{\cal E}_1\!&=\!\frac{1-2y}{\sqrt{1-3y}}\bigg[1-\frac{\bar{u}^\alpha \bar{u}^\beta h_{\alpha\beta}}{2}-\frac{r}{4}\bar{u}_\alpha \bar{u}_\beta \frac{\partial h^{\alpha\beta}}{\partial r}-\frac{r(1-6y)}{12(1-2y)}\bar{u}^\alpha \bar{u}^\beta \frac{\partial h_{\alpha\beta}}{\partial r}\bigg] \, , \\
{\cal J}_1\!&=\!\frac{m_{2}}{\sqrt{y(1-3y)}}\bigg[1-\frac{\bar{u}^\alpha \bar{u}^\beta h_{\alpha\beta}}{2}-\frac{r(1-2y)}{4y}\bar{u}_\alpha \bar{u}_\beta \frac{\partial h^{\alpha\beta}}{\partial r}-\frac{r(1-6y)}{12y}\bar{u}^\alpha \bar{u}^\beta \frac{\partial h_{\alpha\beta}}{\partial r}\bigg] \, ,
	\end{align}
\end{subequations}
which, as demonstrated at the end of Appendix \ref{SFdetails}, are not gauge invariant.  Nevertheless, Eqs.~\eqref{detEJ} can be combined to give 
\beq\label{z1EJ1}
z_1 = {\cal E}_1 - \Omega {\cal J}_1 = \sqrt{1-3y} \, \bigg( 1 - \frac{1}{2} \bar{u}^{\alpha}\bar{u}^{\beta}h_{\alpha\beta} \bigg) \, ,
\eeq
where the term in parenthesis is gauge invariant (see \cite{De.08} and \eqref{eq_uuh} in our Appendix \ref{SFdetails}).

Inspired by the present work, new definitions of energy and angular momentum (per unit mass) of the particle can be given, which are \textit{separately gauge invariant}. Indeed, Eq.~\eqref{partial_m1} provokes a powerful suggestion to introduce these alternative gauge invariant quantities:\footnote{The partial derivatives with respect to the masses are taken with constant orbital frequency $\Omega$. Thus we really mean for instance $\tilde{\cal E}_1=(\partial M/\partial m_1)|_{\Omega, m_2}$.}
\begin{subequations}\label{EJtilde}
	\begin{align}
		 \tilde{\cal E}_1 &\equiv \frac{\partial M}{\partial m_1}\, , \\
		 \tilde{\cal J}_1 &\equiv \frac{\partial J}{\partial m_1}\, .
	\end{align}
\end{subequations}
Note that by definition the same combination as in Eq.~\eqref{z1EJ1} holds, namely
\beq\label{z1EJ1tilde}
z_1=\tilde{\cal E}_1 - \Omega \tilde{\cal J}_1\, .
\eeq
Furthermore, the thermodynamic relation for these quantities holds, \textit{i.e.}
\beq\label{thermo1}
\frac{\partial\tilde{\cal E}_1}{\partial\Omega} = \Omega\,\frac{\partial\tilde{\cal J}_1}{\partial\Omega}\, .
\eeq
This follows from Eq.~\eqref{partial_Omega} and commutation of partial derivatives. Combining Eqs.~\eqref{z1EJ1tilde} and \eqref{thermo1} yields
\begin{subequations}\label{pert_thermo}
	\begin{align}
		\tilde{\cal E}_1&=z_1-\Omega\frac{\partial z_1}{\partial\Omega} \, , \\
		\tilde{\cal J}_1&=-\frac{\partial z_1}{\partial\Omega} \, .
	\end{align}
\end{subequations}
The associated invariant energy and angular momentum are defined by $\tilde{E}_1 \equiv m_1\tilde{\cal E}_1$ and $\tilde{J}_1 \equiv m_1\tilde{\cal J}_1$. Using the first integral relation \eqref{first_integral}, it is straightforward to verify the following connections with the ADM mass and angular momentum:
\begin{subequations}\label{sumEJ}
	\begin{align}
		\tilde{E}_1 + \tilde{E}_2 &= M + \Omega \frac{\partial M}{\partial \Omega}\, , \\
		\tilde{J}_1 + \tilde{J}_2 &= J + \frac{\partial (\Omega J)}{\partial \Omega}\, .
	\end{align}
\end{subequations}

We now compute the gauge invariant quantities \eqref{EJtilde} or \eqref{pert_thermo} to first order in the mass ratio $q$. This readily follows from the expressions \eqref{pert_thermo} which can be rewritten, using the gauge-invariant result for $z_1$ given by \eqref{z1EJ1}, as {(with $r_\Omega=m_2/y$)}
\begin{subequations}\label{EJcal}
	\begin{align}
\tilde{\cal E}_1&=\frac{1-2y}{\sqrt{1-3y}}\bigg[1-\frac{\bar{u}^\alpha \bar{u}^\beta h_{\alpha\beta}}{2}-\frac{r_\Omega(1-3y)}{3(1-2y)}\frac{\partial (\bar{u}^\alpha \bar{u}^\beta h_{\alpha\beta})}{\partial r_\Omega}\Bigg] \, , \label{Ecal} \\
\tilde{\cal J}_1&=\frac{m_{2}}{\sqrt{y(1-3y)}}\Bigg[1-\frac{\bar{u}^\alpha \bar{u}^\beta h_{\alpha\beta}}{2}-\frac{r_\Omega(1-3y)}{3y}\frac{\partial (\bar{u}^\alpha \bar{u}^\beta h_{\alpha\beta})}{\partial r_\Omega}\Bigg] \, ,
	\end{align}
\end{subequations}
where the final results here expressly hold to first perturbative order. These allow us to show, up to terms $\mathcal{O}(q^2)$, that we can write ${\cal E}_1=\tilde{\cal E}_1+\delta {\cal E}$ and ${\cal J}_1=\tilde{\cal J}_1+\delta {\cal J}$, where 
\beq\label{whit_delta}
	\delta {\cal E} = \frac{1-2y}{\sqrt{1-3y}} \bigg[\frac{r_\Omega}{4}\bigg(\frac{\partial (\bar{u}_\alpha \bar{u}_\beta h^{\alpha\beta})}{\partial r_\Omega}-\bar{u}_\alpha \bar{u}_\beta \frac{\partial h^{\alpha\beta}}{\partial r}\bigg)\!+\!\frac{r_\Omega(1-6y)}{12(1-2y)}\bigg(\frac{\partial (\bar{u}^\alpha \bar{u}^\beta h_{\alpha\beta})}{\partial r_\Omega}-\bar{u}^\alpha \bar{u}^\beta \frac{\partial h_{\alpha\beta}}{\partial r}\bigg)\bigg] \, ,
\eeq
together with $\delta {\cal J}=\delta {\cal E}/\Omega$. Note that the terms in the first brackets have different meanings; the first term means the derivative between different circular orbits, and is gauge invariant while, as explained in Appendix \ref{SFdetails}, the second term is not gauge invariant;  rather, as pointed out in Appendix \ref{SFdetails}, it means the derivative of spacetime dependence of $\bar{u}_\alpha \bar{u}_\beta h^{\alpha\beta}$ at the existing quasi-circular orbit with frequency, $\Omega$, held fixed. 

It will be interesting to understand the implications of the new gauge invariant quantities $\tilde{\cal E}_1$ and $\tilde{\cal J}_1$ as we proceed to calculations at second perturbative order. Even without $\tilde{\cal E}_1$ and $\tilde{\cal J}_1$, the relations \eqref{PDE} have another powerful implication, namely that $z_1$ (and hence $u_1^t$) can be calculated directly from the post-Newtonian energy and angular momentum $E$ and $J$ coming from the PN equations of motion, instead of the long PN reduction of the defining expression \eqref{u1t} as was done in Paper I. This thus gives us a new tool through which we will be able to compare future PN results that have been obtained in different ways.

\subsubsection{Post-Newtonian expansions of the invariant energy and angular momentum}
\label{ej_PN}

The new gauge-invariant energy $\tilde{E}_1=m_1\tilde{\cal E}_1$ and angular momentum $\tilde{J}_1=m_1\tilde{\cal J}_1$ of the particle 1 (with smallest mass $m_1$) as defined by \eqref{EJtilde} have been obtained at first order in the mass ratio $q$ in Eqs.~\eqref{EJcal}. This result applies in the strong field regime, and is formally valid up to any post-Newtonian order. We now present the PN expressions of $\tilde{\cal E}_1$ and $\tilde{\cal J}_1$, valid at a finite PN order but for any mass ratio. These can be obtained using Eqs.~\eqref{pert_thermo} and the PN expansion of $z_1$ given by Eq.~\eqref{z1_x}. To write PN expressions valid for an arbitrary mass ratio, it is appropriate to use the variables $x=(m\Omega)^{2/3}$, $\nu=m_1m_2/m^2$, and $\Delta=(m_2-m_1)/m$. We find, up to 3PN order and augmented by the logarithmic contributions at 4PN and 5PN orders, 
\begin{align}
	\tilde{\cal E}_1 &= 1 + \left( - \frac{1}{4} - \frac{1}{4} \Delta + \frac{\nu}{6} \right) x + \left( \frac{3}{16} + \frac{3}{16} \Delta + \frac{\nu}{6} + \frac{1}{24} \Delta \, \nu - \frac{5}{72} \nu^2 \right) x^2 \nonumber \\ &\qquad\! + \left( \frac{27}{32} + \frac{27}{32} \Delta + \frac{\nu}{2} - \frac{19}{16} \Delta \, \nu + \frac{39}{32} \nu^2 + \frac{1}{32} \Delta \, \nu^2 - \frac{\nu^3}{16} \right) x^3 \nonumber \\ &\qquad\! + \left(  \frac{675}{256} + \frac{675}{256} \Delta + \left[ - \frac{190}{9} + \frac{205}{192} \pi^2 \right] \nu + \left[ - \frac{34445}{1152} + \frac{205}{192} \pi^2 \right] \Delta \, \nu \right. \nonumber \\ &\qquad\qquad\!\! \left. + \left[  \frac{19315}{1728} - \frac{205}{576} \pi^2 \right] \nu^2 + \frac{155}{128} \Delta \, \nu^2 - \frac{4865}{2592} \nu^3 + \frac{35}{5184} \Delta \, \nu^3 - \frac{455}{31104} \nu^4 \right) x^4 \nonumber \\ &\qquad\! + \left( \frac{3969}{512} +  \frac{3969}{512} \Delta + \nu \, \biggl[ - \frac{64}{15} -  \frac{64}{15} \Delta - \frac{7}{3} p_4(\nu) - \frac{7}{3} \Delta\,q_4(\nu) \biggr] - \frac{64}{45} \nu^2 \right. \nonumber \\ &\qquad\qquad\!\! \left. + \left[ - \frac{224}{15} - \frac{224}{15} \Delta - \frac{224}{45} \nu \right] \nu \, \ln x \right) x^5 \nonumber \\ &\qquad\! + \left( \frac{45927}{2048} + \, \frac{45927}{2048}\Delta + \nu \, \biggl[ \frac{4988}{315} + \frac{4988}{315} \Delta - 3 p_5(\nu) - 3 \Delta\,q_5(\nu) \biggr] \right. \nonumber \\ &\qquad\qquad\!\! \left. + \,\frac{11876}{315}\nu^2+\frac{328}{15}\Delta\nu^2-\frac{656}{45}\nu^3\right. \nonumber \\ &\qquad\qquad\!\! \left. + \left[  \frac{2494}{35} + \frac{2494}{35} \Delta + \frac{5938}{35} \nu + \frac{492}{5} \Delta \nu - \frac{328}{5} \nu^2 \right] \nu \, \ln x \right) x^6 \, , \label{E1_x}
\end{align}
together with
\begin{align}
	\tilde{\cal J}_1 &= \frac{1}{\Omega} \, \biggl\{ \left( \frac{1}{2} + \frac{1}{2} \Delta - \frac{\nu}{3} \right) x + \left( \frac{3}{4} + \frac{3}{4} \Delta + \frac{2\nu}{3} + \frac{1}{6} \Delta \, \nu - \frac{5}{18} \nu^2 \right) x^2 \nonumber \\ &\qquad\! + \left( \frac{27}{16} + \frac{27}{16} \Delta + \nu - \frac{19}{8} \Delta \, \nu + \frac{39}{16} \nu^2 + \frac{1}{16} \Delta \, \nu^2 - \frac{\nu^3}{8} \right) x^3 \nonumber \\ &\qquad\! + \left(  \frac{135}{32} + \frac{135}{32} \Delta + \left[ - \frac{304}{9} + \frac{41}{24} \pi^2 \right] \nu + \left[ - \frac{6889}{144} + \frac{41}{24} \pi^2 \right] \Delta \, \nu \right. \nonumber \\ &\qquad\qquad\!\! \left. + \left[  \frac{3863}{216} - \frac{41}{72} \pi^2 \right] \nu^2 + \frac{31}{16} \Delta \, \nu^2 - \frac{973}{324} \nu^3 + \frac{7}{648} \Delta \, \nu^3 - \frac{91}{3888} \nu^4 \right) x^4 \nonumber \\ &\qquad\! + \left( \frac{2835}{256} +  \frac{2835}{256} \Delta + \nu \, \biggl[ - \frac{64}{15} - \frac{64}{15} \Delta - \frac{10}{3} p_4(\nu) - \frac{10}{3} \Delta\,q_4(\nu) \biggr] - \frac{64}{45}\nu^2\right. \nonumber \\ &\qquad\qquad\!\! \left. + \left[ - \frac{64}{3} - \frac{64}{3} \Delta - \frac{64}{9} \nu \right] \nu \, \ln x \right) x^5 \nonumber \\ &\qquad\! + \left( \frac{15309}{512} + \, \frac{15309}{512}\Delta + \nu \, \biggl[ \frac{4988}{315} + \frac{4988}{315}\Delta - 4 p_5(\nu) - 4 \Delta\,q_5(\nu) \biggr] \right. \nonumber \\ &\qquad\qquad\!\! \left. + \,\frac{11876}{315}\nu^2 +\frac{328}{15}\Delta\nu^2 -\frac{656}{45}\nu^3\right. \nonumber \\ &\qquad\qquad\!\! \left. + \left[ \frac{9976}{105} + \frac{9976}{105} \Delta + \frac{23752}{105} \nu + \frac{656}{5} \Delta \nu - \frac{1312}{15} \nu^2 \right] \nu \, \ln x \right) x^6\biggr\} \, . \label{J1_x}
\end{align}
The functions $p_4(\nu)$, $p_5(\nu)$ and $q_4(\nu)$, $q_5(\nu)$ have been related to the functions $e_4(\nu)$, $e_5(\nu)$ in the binding energy $E$ by Eqs.~\eqref{pq45}. From these expressions it is easy to verify that $\tilde{\cal E}_1-\Omega\tilde{\cal J}_1=z_1$ holds, together with the thermodynamic relation \eqref{thermo1} {and the connections to the ADM quantities found in Eqs.~\eqref{sumEJ}}.

\subsection{Determination of high order PN coefficients in the binding energy}
\label{E_PN}

Black hole perturbation theory is usually formulated as an expansion in powers of the mass ratio $q = m_1/m_2$. However, at first order in $q$, the symmetric mass ratio $\nu = q / (1+q)^2$ coincides with $q$, \textit{i.e.} $q = \nu + \mathcal{O}(\nu^2)$. In the extreme mass ratio limit $q \ll 1$, we may therefore expand the redshift observable associated with particle $1$ as 
\beq\label{z1exp}
	z_1(x,\nu) = z_\text{Schw}(x) + \nu \, z_\text{SF}(x) + \mathcal{O}(\nu^2) \, ,
\eeq
where we recall that $x = (m\Omega)^{2/3}$. The result for a test particle on a circular orbit around a Schwarzschild black hole is known in closed form as $z_\text{Schw} = \sqrt{1-3x}$; see Eq.~\eqref{z1schw} (for simplicity's sake we remove the label 1 indicating the first particle). The invariant relation $z_\text{SF}(x)$ encoding the first order mass ratio correction was first computed numerically, in the Regge-Wheeler gauge, in Ref.~\cite{De.08}. Alternative self-force (SF) calculations based on different gauges (Lorenz gauge and radiation gauge) were later found to be in agreement within the numerical uncertainties~\cite{Sa.al.08,Sh.al.11}.

The conservative gravitational SF effect $z_\text{SF}(x)$ has also been computed up to high orders in the post-Newtonian approximation. From the general form of the near-zone PN metric, the conservative SF effect on $z_1$ reads
\beq\label{zSF}
	z_\text{SF}(x) = \sum_{k \geqslant 0} \gamma_k \, x^{k+1} + \ln{x} \, \sum_{k \geqslant 4} \delta_k \, x^{k+1} + \cdots \, ,
\eeq
where $k$ is a positive integer, the coefficients $\gamma_k$ and $\delta_k$ are pure numbers, the first logarithms occur at 4PN order, and the dots stand for terms involving higher powers of logarithms $(\ln{x})^p$, with $p \geqslant 2$, which are expected not to occur before the very high 7PN order~\cite{Bl.al2.10}.\footnote{The general structure of the near-zone PN expansion is known to be of the type $\sum x^{n/2} (\ln{x})^p$, where $n$ and $p$ are positive integers~\cite{BlDa.86}. For conservative effects $n/2=k$ is a positive integer.}

The exact values of the Newtonian, 1PN, 2PN and 3PN polynomial coefficients $\gamma_0$, $\gamma_1$, $\gamma_2$ and $\gamma_3$, as well as those of the 4PN and 5PN logarithmic coefficients $\delta_4$ and $\delta_5$ can immediately be derived from the result \eqref{z1_x}, valid for any mass ratio. These analytical results are reported in Table~\ref{tab:analytical_coeffs}.\footnote{Papers I and II actually determined the coefficients $\alpha_k$ and $\beta_k$ in the post-Newtonian expansion in powers of $y = (m_2 \Omega)^{2/3}$ of the self-force effect $u^t_\text{SF}$ on $u_1^t = 1 / z_1$, defined by analogy with Eqs.~\eqref{z1exp} and \eqref{zSF}:
$$u^t_\text{SF}(y) = \sum_{k \geqslant 0} \alpha_k \, y^{k+1} + \ln{y} \, \sum_{k \geqslant 4} \beta_k \, y^{k+1} + \cdots \, .$$
The two sets of coefficients $\{\alpha_k,\beta_k\}$ and $\{\gamma_k,\delta_k\}$ can very easily be related using
$$z_\text{SF}(x) = \frac{x}{\sqrt{1-3x}} - (1-3x) \, u^t_\text{SF}(x) \, .$$
For convenience, we provide the values of both sets of coefficients in Tables \ref{tab:analytical_coeffs} and \ref{tab:numerical_coeffs}.}

\begin{table}[h]
	\begin{tabular}{c|c||c|c}
		\hline\hline
		Coeff. & Value & Coeff. & Value \\
		\hline
		$\alpha_0$ & $-1$ & $\gamma_0$ & $2$ \\
		$\alpha_1$ & $-2$ & $\gamma_1$ & $\frac{1}{2}$ \\
		$\alpha_2$ & $-5$ & $\gamma_2$ & $\frac{19}{8}$ \\
		$\alpha_3$ & $-\frac{121}{3}+\frac{41}{32}\pi^2$ & $\gamma_3$ & $\frac{1621}{48} - \frac{41}{32}\pi^2$ \\
		$\beta_4$ & $-\frac{64}{5}$ & $\delta_4$ & $\frac{64}{5}$ \\
		$\beta_5$ & $\frac{956}{105}$ & $\delta_5$ & $-\frac{4988}{105}$ \\
		\hline\hline
	\end{tabular}
	\caption{The analytically determined PN coefficients $\{\alpha_k,\beta_k\}$ for $u^t_\text{SF}(y)$ and $\{\gamma_k,\delta_k\}$ for $z_\text{SF}(x)$.}
	\label{tab:analytical_coeffs}
\end{table}

Then Paper II showed, for the first time, that it is possible to extract from a SF calculation valuable information corresponding to very high orders in the PN approximation (see also Ref.~\cite{Ba.al.10} for a similar analysis). Indeed, by fitting the highly accurate SF data for $z_\text{SF}(x)$ to a PN series of the type \eqref{zSF}, using the exact values of the analytically determined PN coefficients reported in Table~\ref{tab:analytical_coeffs}, the numerical values of the 4PN, 5PN, and 6PN coefficients $\gamma_4$, $\gamma_5$, $\gamma_6$, and $\delta_6$ could be determined.\footnote{The accuracy of the SF data used in Paper II did not allow an unambiguous distinction between the effects of the 7PN polynomial (\textit{i.e.} $\gamma_7$) and logarithmic contributions ($\delta_7$) in the PN expansion of $z_\text{SF}(x)$. However,  an $\gamma_7$ (and $\alpha_7$) coefficient is included in Table \ref{tab:numerical_coeffs} since it was used in Paper II to ensure the goodness of the fit that was finally obtained for the lower order coefficients. It essentially captures in a single term, to the extent available in the data used, the additional contributions from $\delta_7$ and higher PN order coefficients.  In the remainder of the present work, we shall disregard any results beyond 6PN order, because we believe that a contribution from $\beta_7$ confounds $\alpha_7$ \cite{Bl.al.11}.  We expect the $\alpha_7$ contribution would be substantially improved if PN values could be given for the 6PN and 7PN log coefficients.} These are reported in Table~\ref{tab:numerical_coeffs}. Notice in particular how the 4PN and 5PN coefficients $\gamma_4$ and $\gamma_5$ could be measured with high precision. Note that $\gamma_4$ and $\gamma_5$ coincide with the $\nu \to 0$ limit of the polynomials $p_4(\nu)$, $q_4(\nu)$ and $p_5(\nu)$, $q_5(\nu)$ introduced in Eq.~\eqref{z1_x}, namely
\begin{subequations}\label{pq0}
\begin{align}
\gamma_4 &= p_4(0) + q_4(0) + \frac{1701}{256}\,,\\
\gamma_5 &= p_5(0) + q_5(0) + \frac{15309}{1024} \,.\end{align}
\end{subequations}
\begin{table}[h]
	\begin{tabular}{c|c||c|c}
		\hline\hline
		Coeff. & Value & Coeff. & Value \\
		\hline
		$\alpha_4$ & $-114.34747(5)$ & $\gamma_4$ & $+53.43220(5)$ \\
		$\alpha_5$ & $-245.53(1)$ & $\gamma_5$ & $-37.72(1)$ \\                 
		$\alpha_6$ & $-695(2)$ & $\gamma_6$ & $+123(2)$ \\                 
		$\beta_6$ & $+339.3(5)$ & $\delta_6$ & $-311.9(5)$ \\
		$\alpha_7$ & $-5837(16)$ & $\gamma_7$ & $+4210(9)$ \\
		\hline\hline
	\end{tabular}
	\caption{The numerically determined values of higher order PN coefficients, based on a fit to the SF data given in Paper II. The uncertainties in the last digits are indicated in parenthesis.}
	\label{tab:numerical_coeffs}
\end{table}

Extending to 6PN order the 5PN-accurate expression of the circular-orbit ADM energy $M(\Omega)=m+E(\Omega)$ given by Eq.~\eqref{E_x}, we have
\begin{align}\label{M}
    M &= m - \frac{1}{2} \, m \,\nu \, x \left\{ 1 + \left( - \frac{3}{4} - \frac{\nu}{12} \right) x + \left( - \frac{27}{8} + \frac{19}{8} \nu - \frac{\nu^2}{24} \right) x^2 \right. \nonumber \\ &\qquad\qquad\qquad\quad\, + \left( - \frac{675}{64} + \biggl[ \frac{34445}{576} - \frac{205}{96} \pi^2 \biggr] \nu - \frac{155}{96} \nu^2 - \frac{35}{5184} \nu^3 \right) x^3 \nonumber \\ &\qquad\qquad\qquad\quad\, +\left( - \frac{3969}{128} + \nu \, e_4(\nu) + \frac{448}{15} \nu \ln{x} \right) x^4 \nonumber \\ &\qquad\qquad\qquad\quad\, + \left( - \frac{45927}{512} + \nu \, e_5(\nu) + \biggl[ - \frac{4988}{35} - \frac{656}{5} \nu \biggr] \nu \ln{x} \right) x^5 \nonumber \\ &\qquad\qquad\qquad\quad\, + \left. \left( - \frac{264627}{1024} + \nu \, e_6(\nu) + \nu \, e_6^\text{ln}(\nu) \, \ln{x} \right) x^6 \right\} .
\end{align}
In addition to the 4PN and 5PN unknown coefficients $e_4(\nu)$ and $e_5(\nu)$, we introduced further 6PN unknown coefficients $e_6(\nu)$ and $e_6^\text{ln}(\nu)$, which are also polynomials in $\nu$. We wish to determine the zeroth order coefficients of these four polynomials in the limit $\nu \rightarrow 0$, \textit{i.e.} the numerical values of $e_4(0)$, $e_5(0)$, $e_6(0)$, and $e_6^\text{ln}(0)$. To do so, we shall use the first law, more precisely Eqs.~\eqref{partial_Omega} and \eqref{partial_m1}, together with the SF results for the redshift observable $z_1$.

Since Eq.~\eqref{partial_m1} involves the combination $\mathcal{M} = M - \Omega J$, it will be convenient to express $\mathcal{M}$ as a function of the energy $M$ alone \textit{via} the ``thermodynamic'' relation \eqref{partial_Omega}. Indeed we have $\ud(\mathcal{M}/\Omega)=M\ud(1/\Omega)$, from which we deduce
\beq\label{M_OmegaJ}
	\mathcal{M} = -\Omega \int \frac{M(\Omega)}{\Omega^{2}} \, \ud \Omega = -\frac{3}{2} \, x^{3/2} \int \frac{M(x)}{x^{5/2}} \, \ud x \, ,
\eeq
where $M$ is treated as a function of $\Omega$ or, more precisely, $x$, holding the individual masses $m_1$ and $m_2$ fixed. Introducing the PN expansion \eqref{M} into \eqref{M_OmegaJ} and integrating yields the 6PN-accurate expression\footnote{Alternatively, we could have subtracted directly the PN expansions \eqref{E_x} and \eqref{J_x} and use the relations \eqref{j45} linking the unknown coefficients $j_4(\nu)$ and $j_5(\nu)$ in the angular momentum to $e_4(\nu)$ and $e_5(\nu)$.}
\begin{align}\label{EomegaJ}
    \mathcal{M} &= m - \frac{3}{2} \, m\,\nu \, x \left\{ 1 + \left( \frac{3}{4} + \frac{\nu}{12} \right) x + \left( \frac{9}{8} - \frac{19}{24} \nu + \frac{\nu^2}{72} \right) x^2 \right. \nonumber \\ &\qquad\qquad + \left( \frac{135}{64} + \biggl[ - \frac{6889}{576} + \frac{41}{96} \pi^2 \biggr] \nu + \frac{31}{96} \nu^2 + \frac{7}{5184} \nu^3 \right) x^3 \nonumber \\ &\qquad\qquad + \left( \frac{567}{128} + \biggl[ \frac{128}{105} - \frac{e_4(\nu)}{7} \biggr] \nu - \frac{64}{15} \nu \ln{x} \right) x^4 \nonumber \\ &\qquad\qquad + \left( \frac{5103}{512} + \biggl[ - \frac{9976}{2835} - \frac{e_5(\nu)}{9} - \frac{1312}{405} \nu \biggr] \nu + \biggl[ \frac{4988}{315} + \frac{656}{45} \nu \biggr] \nu \ln{x} \right) x^5 \nonumber \\ &\qquad\qquad + \left( \frac{24057}{1024} + \biggl[ - \frac{e_6(\nu)}{11} + \frac{2}{121} e_6^\text{ln}(\nu) \biggr] \nu - \left. \frac{e_6^\text{ln}(\nu)}{11} \, \nu \ln{x} \right) x^6 \right\} .
\end{align}
In the test-particle limit, we recover the 6PN expansion of the exact result, which reads
\beq
\mathcal{M} = m + m \, \nu \left( \sqrt{1-3 x} - 1 \right) + \mathcal{O}(\nu^2) \, .
\eeq

The relation $z_1=\partial\mathcal{M}/\partial m_1$ [see Eq.~\eqref{partial_m1}] establishes a direct link between the redshift $z_1$ of particle $1$, and the combination $\mathcal{M} = M - \Omega J$ of the ADM energy and angular momentum. Making the change of variables $(\Omega,m_1,m_2) \longrightarrow (\Omega,m,\nu)$, this equation becomes (remember that $m_1 \leqslant m_2$, with $\Delta=(m_2-m_1)/m= \sqrt{1-4\nu}$)
\beq\label{z1M}
	z_1 = \frac{\partial \mathcal{M}}{\partial m} + \frac{1 - 4 \nu + \Delta}{2m} \, \frac{\partial \mathcal{M}}{\partial \nu} \, .
\eeq
The analogous relation for the redshift $z_2$ of particle $2$ is obtained by changing $\Delta$ into $-\Delta$. We then make another change of variables, namely $(\Omega,m,\nu) \longrightarrow (x,m,\nu)$. Since the ratio $\mathcal{M} / m$ does not depend explicitly on the total mass $m$ but only through $x=(m\Omega)^{2/3}$, we have $m\,\partial \mathcal{M} / \partial m = \mathcal{M}+\frac{2}{3}x\,\partial \mathcal{M} / \partial x$; therefore Eq.~\eqref{z1M} can be rewritten as
\beq\label{z1Mbis}
	m \, z_1 = \mathcal{M} + \frac{2x}{3} \, \frac{\partial \mathcal{M}}{\partial x} + \frac{1 - 4 \nu + \Delta}{2} \, \frac{\partial \mathcal{M}}{\partial \nu}\, .
\eeq
We can now expand $\Delta = \sqrt{1-4\nu}$ in powers of $\nu$, and neglect terms $\mathcal{O}(\nu^2)$ because these contributions are not controlled in the self-force calculation of the redshift $z_1$. Equating the terms $\mathcal{O}(\nu)$ in both sides of Eq.~\eqref{z1Mbis}, we obtain the algebraic relations between the known coefficients $\gamma_k$ and $\delta_k$ entering the PN expansion \eqref{zSF} of the SF effect on $z_1$, and the $\nu \to 0$ limit of the unknown coefficients $e_k(\nu)$ and $e_k^\text{ln}(\nu)$ entering the PN expansion \eqref{M} of the energy $M$, which we rearrange as\footnote{Alternatively, the relations \eqref{e4_a4}--\eqref{e5_a5} between $e_4(0)$, $e_5(0)$ and $\gamma_4$, $\gamma_5$ could also be obtained by substituting the $\nu \to 0$ limit of Eqs.~\eqref{pq45} into \eqref{pq0}.}
\begin{subequations}
	\begin{align}
		e_4(0) &= \frac{7}{3} \gamma_4 + \frac{28037}{960} \, , \label{e4_a4} \\
		e_5(0) &= 3 \gamma_5 + \frac{9359293}{161280} \, , \label{e5_a5} \\
		e_6(0) &= \frac{11}{3} \gamma_6 + \frac{2}{3} \delta_6 + \frac{88209}{256} \, , \\
		e_6^\text{ln}(0) &= \frac{11}{3} \delta_6 \, .
	\end{align}
\end{subequations}
Finally, replacing the coefficients $\gamma_4$, $\gamma_5$, $\gamma_6$, and $\delta_6$ by their known numerical values, as given in Table \ref{tab:numerical_coeffs} above, we find 
\begin{subequations}\label{e4_e6}
	\begin{align}
		e_4(0) &= +153.8803(1) \, , \\
		e_5(0) &= -55.13(3) \, , \\
		e_6(0) &= +588(7) \, , \\
		e_6^\text{ln}(0) &= -1144(2) \, .
	\end{align}
\end{subequations}
Any future post-Newtonian calculation of the 4PN, 5PN, or even 6PN dynamics of point-particle binaries will have to be compatible with these numerical values.\footnote{We notice that the numerical values of the coefficients $e_4(0)$ and $e_5(0)$, as predicted in Ref.~\cite{Fa.11} using the recently computed conservative gravitational self-force correction to the Schwarzschild innermost stable circular orbit \cite{BaSa.09}, are off by $\sim 180 \%$ and $\sim 320 \%$, respectively. This probably reflects the fact that, in the extreme mass ratio limit, even a 5PN-accurate formula for the binding energy does not reproduce faithfully the exact relativistic result.} In principle, these could also be recovered from an accurate post-self-force calculation based on second order black hole perturbation theory~\cite{De.12}. We now turn to some possible applications of the numerical results \eqref{e4_e6}.

\subsection{Other applications left for future work}

We have used the first law \eqref{first_law}, together with the recently determined numerical values of high order PN coefficients in the self-force contribution $z_\text{SF}$ to the invariant relation $z_1(\Omega)$ [see Eqs.~\eqref{z1exp}--\eqref{zSF}], to compute new PN coefficients in the binding energy $E(\Omega)$, at leading order beyond the test-particle approximation. However, since the first law has been derived in full general relativity in Sec.~\ref{subsec:proof}, the relations \eqref{partial_Omega}--\eqref{PDE} are expected to hold at all (conservative) orders in a PN expansion. Making use of these relations, together with the numerical results for the exact conservative SF effect $z_\text{SF}(\Omega)$ on the redshift observable \cite{De.08,Sa.al.08,Sh.al.11}, one could in principle compute the fully relativistic SF contribution $E_\text{SF}(\Omega)$ to the binding energy of non-spinning compact binaries \cite{Le.al2.12}.\footnote{Note that in order to do so, it must be assumed that the results of a SF calculation, which involve \textit{one} point particle orbiting a Schwarzschild black hole, can be used in conjunction with the first law \eqref{first_law}, which was itself derived for \textit{two} point masses in the context of PN spacetimes. From this perspective, SF calculations would essentially be treated as post-Newtonian calculations formally including \textit{all} the PN corrections (at linear order in the mass ratio).} This prospect opens up the following applications.

\subsubsection{ISCO shift induced by the conservative piece of the gravitational self-force} 

In the test-particle limit, the innermost stable circular orbit (ISCO) of the Schwarzschild geometry is defined as the point of onset of a dynamical instability for circular orbits (the circular orbit for which the radial frequency squared of an infinitesimal eccentricity perturbation turns negative). The orbital frequency of the Schwarzschild ISCO naturally coincides with the frequency obtained by minimizing the specific energy $E_\text{Schw} = (1-2x)/\sqrt{1-3x} - 1$ of a test mass in circular orbit around a non-rotating black hole. Going beyond the test-particle approximation, the shift of the Schwarzschild ISCO frequency induced by the conservative part of the gravitational self-force has recently been computed~\cite{BaSa.09,BaSa.10}. This genuine strong field result has been used extensively as a reference point for comparison with other analytical and numerical methods.

In a post-Newtonian context, one can also compute an ISCO (for any mass ratio) from a stability analysis of the conservative part of the PN equations of motion \cite{BlIy.03}. Alternatively, the innermost circular orbit (ICO), or minimum energy circular orbit (MECO), is defined as the minimum of the PN binding energy $E(x)$, when it exists \cite{Bl.02}. An extensive comparison of the SF-induced ISCO shift to the numerous PN-based estimates of ISCO and ICO available in the literature was performed in Ref.~\cite{Fa.11}. In particular, the standard Taylor-expanded 3PN result \cite{BlIy.03}, based on a stability analysis criterion of the 3PN equations of motion was shown to perform astonishingly well.

Now, it has been shown on very general ground that, for arbitrary mass ratio compact binaries, the definitions of ISCO and ICO are formally equivalent \cite{Bu.al.03}; this conclusion does not rely on any PN expansion, and only requires that the conservative dynamics of the binary system derives from a Hamiltonian.\footnote{In a post-Newtonian context however, the location of the ICO needs not agree with that of the ISCO, because of the truncation at a finite PN order of the equations defining these notions \cite{BlIy.03,Sc.11}.} Hence, the exact value of the ISCO frequency shift induced by the conservative SF could (in principle) be recovered by minimizing the reduced binding energy $E / \mu = E_\text{Schw}(x) + \nu \, E_\text{SF}(x) + \mathcal{O}(\nu^2)$ \cite{Le.al2.12}.

\subsubsection{Calibration of the potentials entering the effective-one-body metric}

Within the effective-one-body (EOB) framework, the circular-orbit binding energy $E(x)$ is in one-to-one correspondance with the ``temporal'' potential $A \equiv - g_{tt}^\text{eff}$ entering the EOB effective metric. The knowledge of the self-force correction $E_\text{SF}(x)$ to the test-particle result $E_\text{Schw}(x)$ thus immediately translates into the knowledge of the coefficient linear in $\nu$ in the potential $A$,\footnote{Assuming that the usual mapping between the effective and EOB Hamiltonians holds at all PN orders, and that the non-geodesic terms occuring in the expression of the effective Hamiltonian beyond 2PN order are proportional to the radial momentum $p_r$, at all PN orders, thus vanishing for circular orbits. See Refs.~\cite{Da.10,Ba.al.12} for more details.} \textit{i.e.} the function $A_\text{SF}(u)$ such that $A = 1 - 2 u + \nu \, A_\text{SF}(u) + \mathcal{O}(\nu^2)$, where $u = m / r$ is the usual inverse EOB Schwarzschild-like radial coordinate.

Furthermore, by combining this result to the recent constraint obtained from a SF/EOB comparison of the periastron advance of black hole binaries on circular orbits~\cite{Da.10,Ba.al.10}, one could also compute the SF coefficient entering the expression of the ``radial'' EOB potential $B \equiv g_{rr}^\text{eff}$, \textit{i.e.} the function $B_\text{SF}(u)$ such that $B^{-1} = 1 - 2u + \nu \, B_\text{SF}(u) + \mathcal{O}(\nu^2)$. This would complete the determination of the two potentials entering the  EOB effective metric at linear order in the symmetric mass ratio $\nu$. Such results would obviously be very useful for improving the calibration of EOB models \cite{Ba.al.12}.

\subsubsection{Comparison with sequences of quasi-circular initial data in numerical relativity}

The resulting expression for the binding energy $E(x)$ could also be used to revisit comparisons with sequences of quasi-circular initial data in numerical relativity.\footnote{Incidentally, we note that the sequence constructed in Ref.~\cite{Gr.al.02} relies on the first law \eqref{Smarr_2BH_delta}.} For comparable mass black hole binaries, previous comparisons suggested that the convergence of the PN series may improve with respect to the extreme mass ratio limit~\cite{Bl.02,Bl.03,YuBe.08,Le.al.11}. This could be investigated using the fully relativistic result for $E(x)$, keeping in mind that only the first order correction in $\nu$ beyond the test-particle result would be under control.

However, higher-order uncontrolled terms $\mathcal{O}(\nu^n)$ with $n \geqslant 2$ in $E / \mu$ may give only a very small contribution to the exact result. Indeed, in a PN expansion, one can check that the terms $\mathcal{O}(\nu^2)$ and $\mathcal{O}(\nu^3)$ in Eq.~\eqref{M} contribute less than $1\%$ to the total 3PN result, up to the Schwarzschild ISCO at $x = 1/6$. Furthermore, a similar observation has recently been put forward in Ref.~\cite{Le.al.11}, for another coordinate invariant relation, namely the general relativistic periastron advance $\Delta \Phi (x)$, in the case of non-spinning binary black holes on quasi-circular orbits, even in the strong field regime accessible to numerical relativity simulations.

\section*{Acknowledgements}
The authors gratefully acknowledge S. Detweiler for illuminating discussions, and especially for providing some key input at an early stage of this work. We thank L.~Barack, E.~Barausse, A.~Buonanno, J.~Friedman, and E.~Poisson for helpful discussions and/or comments. A.L.T. acknowledges support from NSF Grant No. PHY-0903631 and the Maryland Center for Fundamental Physics, and the hospitality of the Department of Physics at University of Florida, where part of this work was completed. B.F.W. acknowledges support from NSF Grant No. PHY-0855503 and hospitality at the Institut d'Astrophysique de Paris during numerous visits throughout the course of this work. L.B. acknowledges fruitful visits to the University of Maryland and the Rochester Institute of Technology.

\appendix

\section{Details of the self-force analysis} \label{SFdetails}

\subsection{Some relevant definitions} \label{SFdefinitions}

In Ref.~\cite{De.08}, Detweiler defines the quantities ${\cal E}_1$, $\dot {\cal R}_1$, and ${\cal J}_1$ in terms of the particle's four-velocity  and four-momentum by
\begin{equation}
   {\cal E}_1\equiv -u_{1 t} \, , \qquad \rdot \equiv u^r_{1} \, , \qquad\text{and}\qquad  {\cal J}_1 \equiv u_{1 \phi} \, .
\label{utpR}
\end{equation}
In general, they are functions of the proper time, $s$, and the overdot represents $\ud/\ud s$.  In consequence,
\begin{subequations}\label{ueq}
	\begin{align}
		 {u_1}_\alpha &= \biggl( -{\cal E}_1, \frac{\dot {\cal R}_1+ u_{1}^\beta h_\beta{}^r}{1-2 m_2/r} , 0, {\cal J}_1 \biggr) \, , \label{ua} \\
		 u_1^\alpha &= \biggl( \frac{{\cal E}_1+ u_{1}^\beta h_{\beta t}}{1-2 m_2/r}, \dot {\cal R}_1, 0, \frac{{\cal J}_1-u_{1}^\beta h_{\beta\phi}}{{r}^2} \biggr) \, . \label{uA}
	\end{align}
\end{subequations}
For quasi-circular orbits, Detweiler assumes that ${\cal E}_1$, ${\cal R}_1$ and ${\cal J}_1$ all change slowly, so that 
\begin{equation}\label{eqdot}
\dot {\cal E}_1 \sim  \rdot \sim \dot {\cal J}_1 \sim {\cal O}(q)\, ,
\end{equation}
while their rates of change vary more slowly still, so that
\begin{equation}\label{eqddot}
\ddot {\cal E}_1\sim \rddot \sim \ddot {\cal J}_1 \sim {\cal O}(q^2)\, .
\end{equation}
Although not entirely necessary, Detweiler found it convenient to define
\begin{equation}
\ubar^\alpha \equiv {\ubar_1^\alpha =} \left( \frac{{\cal E}_1}{1-2{m_2} /{r}},0, 0,
                  \frac{{\cal J}_1}{{r}^2}   \right) ,
\label{ubar}
\end{equation}
being the non-radial part of the four-velocity of the particle, and with all $h_{\alpha\beta}$ terms removed {(\textit{i.e.} it corresponds to the non-radial velocity in the background)}.

There are two extra quantities that are needed in Sec.~\ref{subsec:newgaugeinv} and which Detweiler introduced in his Appendix A of \cite{De.08}. The first is simply the contraction
\begin{equation}\label{det_uuh}
\ubar^\alpha\ubar^\beta h_{\alpha\beta}=\frac{{\cal E}_1}{1-2 m_2/r}\ubar^\beta h_{t\beta}+\frac{{\cal J}_1}{r^2}\ubar^\beta h_{\phi \beta}\,,
\end{equation}
that follows directly from Eq.~\eqref{ubar}, and the second, in which $\Omega$ is held fixed, is
\begin{equation}
\frac{{\partial}g^{\alpha\gamma}}{{\partial}r}\ubar_\gamma\ubar^\beta h_{\alpha\beta}=-\frac{2m_2\,{\cal E}_1}{(r-2 m_2)^2}\ubar^\beta h_{t\beta}-\frac{2{\cal J}_1}{r^3}\ubar^\beta h_{\phi \beta}\,,
\end{equation}
in which we depart slightly from \cite{De.08}, to emphasize that the derivatives do {\textit{not}} act on ${\cal E}_1$ or ${\cal J}_1$ which, as indicated below \eqref{utpR}, are functions of the proper time, $s$, along the orbit; rather the derivative is simply acting on components of the metric embedded in the definition \eqref{ubar}.

\subsection{Gauge transformation properties} \label{GTdetails}

In the application discussed in Sec.~\ref{subsec:newgaugeinv}, we are interested in identifying gauge invariants.  Here we discuss the transformations of $\ubar^\alpha\ubar^\beta h_{\alpha\beta}$, and related quantities, under infinitesimal gauge transformations generated by a gauge vector $\xi^\alpha$, which leads to a metric perturbation given by
\begin{equation}
\Delta h_{\alpha\beta}=-\nabla_\alpha\xi_\beta-\nabla_\beta\xi_\alpha\, .
\end{equation}
In his analysis \cite{De.08}, Detweiler assumed equatorial symmetry, so that $\xi^\theta$ and all its derivatives vanish on the equatorial plane, and the resulting metric perturbations are given in his Eqs.~(B2)--(B7).  As indicated in Sec.~\ref{subsec:motivation}, and emphasized in Sec.~\ref{sec:log_contrib}, for quasi-circular orbits we require the gauge transformation to obey the implied Killing symmetry, so that we must have $(\partial_t+\Omega\partial_{\phi})\,\xi_{\alpha}=0$.  Making use of Detweiler's (B4)--(B6) and the HKV condition, and discarding quantities higher than first order, we find
\begin{align}
\bar{u}^{\alpha}\bar{u}^{\beta}\Delta h_{\alpha \beta}\equiv&\ \bar{u}_{\alpha}\bar{u}_{\beta}\Delta h^{\alpha \beta}={{\cal E}_1^2\over (1-2m_2/r)^2}\bigg[{2m_2\over r^2}\xi^r+2\Omega\partial_\phi\xi_t\bigg]\cr
&-{2{\cal E}_1\over (1-2m_2/r)}{{\cal J}_1\over r^2}\bigg[\partial_\phi\xi_t+\partial_t\xi_\phi\bigg]+{{\cal J}_1^2\over r^4}\bigg[-2r\xi^r+{2\over \Omega}\partial_t\xi_\phi\bigg]\,,
\end{align}
where the large square brackets contain only metric component transformations.  We can write this alternatively as
\begin{align}\label{eq_uuh}
\bar{u}^{\alpha}\bar{u}^{\beta}\Delta h_{\alpha \beta}&={2{\cal E}_1^2\over (1-2m_2/r)^2}\bigg[r\bigg({m_2\over r^3}-\Omega^2\bigg)\xi^r\bigg]\cr
&+{2{\cal E}_1\over 1-2m_2/r}\bigg({\Omega {\cal E}_1\over 1-2m_2/r}-{{\cal J}_1\over r^2}\bigg)\bigg[\Omega r\xi^r+\partial_\phi\xi_t\bigg]\cr
&-{2{\cal J}_1\over r^2}\bigg({\Omega {\cal E}_1\over 1-2m_2/r}-{{\cal J}_1\over r^2}\bigg)\bigg[-r\xi^r+{1\over \Omega}\partial_t\xi_\phi\bigg]\,,
\end{align}
where, in addition, the large round brackets contain terms which vanish to ${\cal O}(q)$ at the particle. Thus, it is clear that $\bar{u}^{\alpha}\bar{u}^{\beta}\Delta h_{\alpha \beta}$ is gauge invariant when evaluated at the particle. This result is needed in \eqref{z1EJ1} to establish that $z_1$ is gauge invariant through first order in $q$.

To evaluate $\bar{u}^{\alpha}\bar{u}^{\beta}\partial_r\Delta h_{\alpha\beta}$ at the particle, we need differentiate only the first large round bracket, and we obtain the equivalent of Detweiler's equations (B14) and (B15):
\begin{subequations}\label{uuDrh}
\begin{align}
\bar{u}^{\alpha}\bar{u}^{\beta}{\partial\Delta h_{\alpha \beta}\over \partial r}=&-{6{\cal E}_1^2\over (1-2m_2/r)^2}{m_2\over r^3}\xi^r,\quad\hbox{\rm so that}\\
\bar{u}^{\alpha}\bar{u}^{\beta}{\partial\Delta h_{\alpha \beta}\over \partial r}\bigg|_\bullet=&-{6\Omega^2\over (1-3y)}\xi^r \, ,
\end{align}
\end{subequations}
where the notation $|_\bullet$ indicates that the RHS expression has been evaluated {\textit{at}} the particle. This was used by Detweiler to show that $\Omega$ [see his Eq.~(28)] was indeed gauge invariant.

To evaluate $\bar{u}_{\alpha}\bar{u}_{\beta}\partial_r\Delta h^{\alpha \beta}$, we should differentiate all large round brackets, and we find
\begin{subequations}\label{Druuh}
\begin{align}
\bar{u}_{\alpha}\bar{u}_{\beta}\frac{\partial\Delta h^{\alpha \beta}}{\partial r}
&=-6{{\cal E}_1^2\over (1-2m_2/r)^2}\Omega^2\xi^r+6{{\cal E}_1^2\over (1-2m_2/r)^2}\bigg(\Omega^2-{m_2\over r^3}\bigg)\xi^r\cr
&+{4\over r}{(1-3m_2/r)\over (1-2m_2/r)}{\Omega {\cal E}_1^2\over (1-2m_2/r)^2}\bigg[\Omega r\xi^r+\partial_\phi\xi_t\bigg]\cr
&-{4\over r}{{\cal E}_1\over (1-2m_2/r)}\bigg({\Omega {\cal E}_1\over 1-2m_2/r}-{{\cal J}_1\over r^2}\bigg)\bigg[\Omega r\xi^r+\partial_\phi\xi_t\bigg]\\
&-{4\over r}{(1-3m_2/r)\over (1-2m_2/r)}{\Omega^2 {\cal E}_1^2\over (1-2m_2/r)^2}\bigg[-r\xi^r+{1\over \Omega}\partial_t\xi_\phi\bigg]\cr
&+{4\over r}{\Omega {\cal E}_1\over (1-2m_2/r)}\bigg({\Omega {\cal E}_1\over 1-2m_2/r}-{{\cal J}_1\over r^2}\bigg)\bigg[-r\xi^r+{1\over \Omega}\partial_t\xi_\phi\bigg], \quad\hbox{\rm so that}\cr
\bar{u}_{\alpha}\bar{u}_{\beta}\frac{\partial\Delta h^{\alpha \beta}}{\partial r}
\bigg|_\bullet&=\ 2{(1-6y)\over (1-2y)}{\Omega^2\over (1-3y)}\xi^r+{4\over r}{\Omega\over (1-2y)}\bigg[\partial_\phi\xi_t-\partial_t\xi_\phi\bigg]\,,
\end{align}
\end{subequations}
a result not given explicitly in \cite{De.08}. Notice that neither $\bar{u}^{\alpha}\bar{u}^{\beta}\partial_r h_{\alpha \beta}$ nor $\bar{u}_{\alpha}\bar{u}_{\beta}\partial_r h^{\alpha \beta}$ is gauge invariant, even though $\bar{u}^{\alpha}\bar{u}^{\beta}h_{\alpha \beta}\!\equiv\!\bar{u}_{\alpha}\bar{u}_{\beta}h^{\alpha \beta}$ itself is.  The terms discussed in \eqref{uuDrh} and \eqref{Druuh} both occur in \eqref{detEJ}.  We see that under gauge transformations, ${\cal E}_1$ and ${\cal J}_1$ transform according to  $\Delta {\cal E}_1=-\Omega/\sqrt{1-3y}\times[\partial_\phi\xi_t-\partial_t\xi_\phi]$ and $\Delta {\cal J}_1=-1/\sqrt{1-3y}\times[\partial_\phi\xi_t-\partial_t\xi_\phi]$, respectively. They are thus radially gauge invariant, but not under Killing-compatible $t$ and $\phi$ changes.

\bibliography{/home/letiec/Publications/ListeRef}

\end{document}